\documentclass[twocolumn,aps,prc,superscriptaddress,showpacs,floatfix]{revtex4}

\usepackage{epsfig}
\usepackage{amsmath}
\usepackage{multirow}
\usepackage{graphicx}
\usepackage{amsmath}
\usepackage{feynmf}
\usepackage[normalem]{ulem}  % \sout{old text} for strikeout
\usepackage[dvips]{color} % For blue in-text comments and additions

\renewcommand\sout{\bgroup \color{red} \ULdepth=-.5ex \ULset}

\newcommand{\Ex}[2]{\ifmmode{#1\times10^{#2}}\else{$#1\times10^{#2}$}\fi}
\begin{document}
\title{Nuclear Symmetry Energy from QCD sum rules}

\author{Kie Sang Jeong}\affiliation{Institute of Physics and Applied Physics, Yonsei
University, Seoul 120-749, Korea}
\author{Su Houng Lee}\affiliation{Institute of Physics and Applied Physics, Yonsei University, Seoul 120-749, Korea}

\date{\today}
\begin{abstract}
We calculate the nucleon self-energies in isospin-asymmetric nuclear
matter using QCD sum rules. Taking the difference of these for the
neutron and proton enables us to express the potential part of the
nuclear symmetry energy in terms of local operators. We find that
the scalar (vector) self energy part gives a negative (positive)
contribution to the nuclear symmetry energy which is consistent with
the results from relativistic mean field theories. Moreover, we find
that an important contribution to the negative contribution of the
scalar self energy comes from the twist-4 matrix elements, whose
leading density dependence can be extracted from deep inelastic
scattering experiments. This suggests that the twist-4 contribution
partly mimics the exchange of the $\delta$ meson and that it
constitutes an essential part in the origin of the nuclear symmetry
energy from QCD. Our result also extends an early success of QCD sum
rule method in understanding the symmetric nuclear matter in terms
of QCD variables to the asymmetric nuclear matter case.
\end{abstract}
\maketitle

\section{Introduction}

There is a renewed interest in the study of nuclear symmetry energy
recently,  as  the next generation low-energy rare isotope
accelerators are being constructed and planned
worldwide~\cite{Li:2008gp}.   Understanding the details of the
nuclear symmetry energy is intricately related to understanding a
wide range of subjects ranging from rare isotopes to neutron rich
nuclear matter such as the neutron star~\cite{Xu:2010xh,Li:2011xr}.
One of the main puzzle to be solved currently is the behavior of the
nuclear symmetry energy at high
density~\cite{Wen:2009av,Xiao:2009zza}.

From a phenomenological point of view, the nuclear symmetry energy
can be obtained by looking at  the nuclear binding energy within the
semi-empirical mass formula in the limit of a large nucleon
number~\cite{Chen:2011ek}. There, the symmetry energy can be
understood as originating from the energy difference between the
proton and the neutron in isospin asymmetric nuclear matter. Hence,
in this picture, the nuclear symmetry energy can be obtained from
the nucleon optical potential or by calculating the energy of the
nucleon quasi-particle near the Fermi surfaces in asymmetric nuclear
matter.

In the  Dirac phenomenology of nucleon-nucleus
scattering~\cite{Wallace,Serot:1984ey}, the optical potential of the
nucleon is composed of a vector and scalar part, $U \simeq
S+V\gamma^0$.  It is well known that in order to fit the spin
observable, one needs a strong scalar  attraction (Re $S<0$) and a
strong vector repulsion (Re $V>0$) both of several hundred MeV, but
such that the combined sum to the energy is only a few tens of MeV,
a result consistent with traditional low-energy nuclear physics. The
strong scalar and vector potentials appear naturally in the
relativistic mean-field theories (RMFT), where meson exchange
interactions between nucleons on the Fermi sea produces the strong
scalar and vector potentials for the nucleons.

But it was only after the works in QCD sum rules that the strong
optical potentials were found to have a basis in QCD. The
application of QCD sum rules \cite{Shifman:1978bx,Reinders:1984sr}
to the nucleon in the vacuum was  developed in
Ref.~\cite{Ioffe:1981kw,Chung}.   The first pioneering work of
applying the QCD sum rule method to nucleons in medium was performed
by Drukarev and Levin~\cite{Drukarev:1988kd,Drukarev:1991fs}.  Here,
the operator product expansion (OPE) was performed in the light cone
direction where $-q^2, q \cdot u \rightarrow \infty$ with their
ratio finite, where $q^\mu,u^\mu$ are the external momenta and the
medium four vector respectively. Later, the relation became clearer
through the work by Cohen, Furnstahl and
Griegel~\cite{Cohen:1991js}, who showed that the strong
scalar-vector self energy appearing in the quasi nucleon pole in the
symmetric nuclear matter can be traced back to the scalar-vector
quark condensate in the nuclear medium. The OPE in this work was
based on the short distance expansion, where  $-q^2 \rightarrow
\infty$, while $qu $ is held fixed. For the medium at rest, this
expansion is equivalent to taking the energy to be large and
imaginary at a fixed finite three momentum~\cite{Furnstahl:1992pi,
Jin:1992id,Jin:1993up,Cohen:1994wm} and, hence, the comparison to
the self-energy obtained in the RMFT approaches becomes more direct
through the use of the energy dispersion relation. \footnote{ One
should caution, however, that the strong medium dependence of the
scalar-scalar four-quark condensate obtained from a naive factorized
form leads to a result that does not agree well with the nuclear
phenomenology~\cite{Jin:1993up}.}

Motivated by these results, and to express and elucidate the origin
of nuclear symmetry energy directly from QCD, we have applied the
QCD sum rule to calculate the neutron and proton energy in
asymmetric nuclear matter. Identifying the difference with
appropriate factors to the nuclear symmetry energy, we show that
this energy can be expressed in terms of quark and gluon degrees of
freedom. Results based on the first formalism to calculate the
nucleon mass in asymmetric matter using QCD sum rules were reported
before~\cite{Drukarev:2004fn,Drukarev:2010xv,Kryshen:2011ng}.
 But here, we will follow the second
formalism adopted in Ref.~\cite{Cohen:1991js}. We have performed the
OPE up to dimension-six operators and have identified all the
independent twist-4 operators. Independent twist-4 operators and
their relation to moments of structure functions appearing in deep
inelastic scattering (DIS) were identified before
\cite{Politzer,Shuryak1,Shuryak2,Jaffe:1981td,Jaffe:1982pm,Ellis1,Ellis2}.
In a later work by one of us \cite{Choi:1993cu,Lee:1993ww}, the
available experimental data on twist-4 effects were collected to
constrain the independent matrix elements.  Using this information,
we have calculated the leading density dependence on the nucleon sum
rules coming  from the  twist-4 effects. From the QCD sum rule
analysis, we find that the scalar (vector) self energy part gives a
negative (positive) contribution to the nuclear symmetry energy,
which is consistent with the results from relativistic mean-field
theories. Moreover, we find that an important contribution to the
negative contribution of the scalar self-energy comes from the
twist-4 matrix elements, whose higher-density behavior will
determine the still-controversial property of the symmetry energy at
these densities.  Our result also extends an early success of the
QCD sum rule method in understanding the symmetric nuclear matter in
terms of QCD operators to the asymmetric nuclear matter case.

The paper is organized as follows. In Sec.~II, we start with a brief
review and a simple idea for the nuclear symmetry energy. In
Sec.~III, we develop the  QCD sum rule formalism, and  discuss the
OPE and its matrix elements. The results for the QCD sum rule
analysis for the nucleons in asymmetric nuclear matter and the
nuclear symmetry energy are presented in Sec.~IV.  Finally, the
conclusion is given in Sec.~V.

\section{A simplified description for The Nuclear Symmetry Energy}

We start from a finite nuclei with $A$ nucleons.
The Bethe-Weizs\"{a}ker formula for the nuclear binding energy is given as
\begin{align}
m_{\textrm{tot}}&=Nm_n+Zm_p-E_B/c^2,\nonumber\\
E_B&=a_VA-a_SA^\frac{2}{3}-a_C(Z(Z-1))A^{-\frac{1}{3}}\nonumber\\
&\quad-a_AI^2A+\delta(A,Z),\label{semp}
\end{align}
where $I = (N-Z)/A$.  The fourth term accounts for the total shifted
energy due to the neutron number excess.  Taking the infinite nuclear
matter limit of this formula, one notes that $a_A$ reduces to the
nuclear symmetry energy~\cite{Chen:2011ek}.

To derive the formula for $a_A$ that can be generalized to the
infinite nuclear matter, we start from a simple formula for the
total energy:
\begin{align}
E_{\textrm{tot}} & = N \overline{E}_n + Z \overline{E}_p \nonumber \\
& =  \frac{1}{2}A(1+I) \overline{E}_n + \frac{1}{2}A(1-I) \overline{E}_p  \nonumber \\
& =  \frac{1}{2}A( \overline{E}_n +  \overline{E}_p) +\frac{1}{2}A
I( \overline{E}_n -\overline{E}_p),\label{define}
\end{align}
where $\overline{E}_n$ ($\overline{E}_p$) is the average neutron
(proton) quasi-particle energy in asymmetric nuclear matter. Now,
the core of the model is what approximation goes into calculating
the average energy.

The symmetry energy in asymmetric nuclear matter is defined as
\begin{align}
E_{\textrm{tot}}(\rho,I)&=E_0(\rho)A+E_{\textrm{sym}}(\rho)I^2A+O(I^4),
\label{def-sym}
\end{align}
where $\rho$ is the nuclear medium density and $I = (N-Z)/A
\rightarrow (\rho_n-\rho_p)/(\rho_n+\rho_p)$ and the neutron and proton
densities are $\rho_n = \frac{1}{2} \rho(1+ I)$, $\rho_p = \frac{1}{2}
\rho(1- I)$ respectively. Therefore, in Eq.~\eqref{define}, the
symmetry energy will have contributions from the term proportional
to $I$ in $( \overline{E}_n -  \overline{E}_p)$ and the term
proportional to $I^2$ in $( \overline{E}_n + \overline{E}_p)$.

For a non interacting Fermi gas of nucleons, each with mass $m_N$,
calculating the average nucleon energy will give
$\overline{E}=\frac{3}{5} E_F$, where $E_F$ is the nucleon Fermi
energy. Following the procedure described above and extracting the
term proportional to $I^2$ gives a nuclear symmetry energy of
$\frac{1}{3} E_F$.

Going back to finite nuclei, assuming a `Fermi well' with constant
energy difference $\Delta$ between adjacent nucleon energy levels,
the symmetry energy can be obtained from the second term of
Eq.~\eqref{define}.  That is, using $ ( \overline{E}_n - \overline{E}_p) =
\frac{1}{4}IA \Delta $, we have,
\begin{align}
a_A=& \frac{1}{8}A \Delta =\frac{1}{4I}(E_n(A,I)-E_p(A,I)).
\label{sym1}
\end{align}

For the infinite nuclear matter case, we can calculate $\overline{E}_N$ from
\begin{align}
\overline{E}_N  = & \frac{1}{\int d^3 k_n d^3 k_p} \int d^3 k_n d^3
k_p E_N(\rho_n,\rho_p),
\end{align}
and obtain  the nuclear symmetry energy $E^{\textrm{sym}}(\rho)$, as
it appears in Eq.~\eqref{def-sym}, by collecting coefficients of
$I^2$ in Eq.~\eqref{define}.  $E^{\textrm{sym}}(\rho)$ can in
general be decomposed into the kinetic like part and the potential
like part in the mean field type quasi-particle approximation. The
kinetic part of $E^{\textrm{sym}}$ can be obtained from the formula
given in Ref.~\cite{Kubis:1997ew};
\begin{align}
E^{\textrm{sym}}_{K}=\frac{1}{6}\frac{k_F^2}{\sqrt{k_F^2+E_{q,V(I=0)}^{2}}},\label{kinetic}
\end{align}
where $k_F$ is the Fermi momentum and $E_{q,V(I=0)}$ is the
potential part of the quasi-nucleon self energy in a symmetric nuclear
matter.

%Brief idea for finding the potential part follows.

\subsection{Linear density approximation}

In the present QCD sum rule calculations, we will be using the
linear density approximation, because the in-medium condensates in
the QCD sum rule can be most reliably estimated to leading order in
density.  This means that for either the proton or the neutron, the
mass will be given as follows:
\begin{align}
E^n_{V}(\rho_n,\rho_p) & =  m_0+a \rho_p+b \rho_n  \nonumber \\
        & =  m_0+\frac{1}{2} \rho(a+b) +\frac{1}{2}  \rho I(b-a), \nonumber \\
E^p_{V}(\rho_n,\rho_p) & =  m_0+\frac{1}{2} \rho(a+b)
-\frac{1}{2}\rho I (b-a),
\end{align}
where $m_0$ is the vacuum mass and $a,b$ are the constants to be determined
later. We then have,
\begin{align}
\overline{E}^N_V & =  \frac{1}{\int d^3 k_n d^3 k_p} \int d^3 k_n d^3 k_p E^N_V(\rho_n,\rho_p)  \nonumber \\
 &=   m_0+ \frac{1}{2} a \rho_p+ \frac{1}{2} b \rho_n  \nonumber \\
 &=   m_0+ \frac{1}{4}\rho (a+b) + \frac{1}{4} \rho I(b-a). \label{define2}
\end{align}
Combining Eq.~\eqref{define2} with Eq.~\eqref{define}, we obtain the
symmetry energy. That is, $ ( \overline{E}^n_{V} -
\overline{E}^p_{V}) = \frac{1}{2} [E^n_{V}(\rho_n, \rho_p) -
E^p_{V}(\rho_n,\rho_p)] $, hence,
\begin{align}
E^{\textrm{sym}}_{V}& =\frac{1}{4I}( E^n_{V}(\rho_n, \rho_p) -
E^p_{V}(\rho_n,\rho_p)),\label{sym2}
\end{align}
which is similar to the relation given in Eq.~\eqref{sym1}. Therefore, to
this order, the symmetry energy comes only from the energy
difference in the proton and neutron at the Fermi surface in
an asymmetric nuclear matter as given in Eq.~\eqref{sym2}. However,
when operators have higher density dependence,
the factors appearing in Eq.~\eqref{define2} should be modified, and the
symmetry energy will have contributions from both the sum and the
difference of the nucleon energies.

The quantity of interest, namely $[E^n_{V}(\rho_n, \rho_p) -
E^p_{V}(\rho_n,\rho_p)]$ can be obtained by looking at the pole of
the nucleon propagator in nuclear medium:
\begin{align}
G(q) = -i \int d^4 x e^{iqx} \langle\Psi_0\vert \textrm{T}[\psi(x)
\bar{\psi}(0) ] \vert\Psi_0\rangle \label{4},
\end{align}
where $\vert\Psi_0\rangle$ is the nuclear medium ground state, and
$\psi(x)$ is a nucleon field. A relativistic mean-field type of
contribution will then appear in the self-energies. The nucleon
propagator can be decomposed as
\begin{align}
G(q) = G_s(q^2,qu)+G_q(q^2,qu) \slash \hspace{-0.2cm} q+G_u(q^2,qu)
\slash \hspace{-0.2cm}u,
\end{align}
where $u^\mu$ is the four-velocity of the nuclear medium ground
state~\cite{Furnstahl:1992pi}.

The nucleon self energy can be
decomposed similarly as~\cite{Cohen:1991js, Furnstahl:1992pi,
Jin:1992id, Jin:1993up}:
\begin{align}
\Sigma(q)&=\tilde{\Sigma}_s(q^2,q
u)+\tilde{\Sigma}_v^\mu(q)\gamma_\mu,
\end{align}
where
\begin{align}
\tilde{\Sigma}_v^\mu(q)&=\Sigma_u(q^2,q
u)u^\mu+\Sigma_q(q^2,qu)q^\mu.
\end{align}
In the mean-field approximation $\Sigma_s$ and $\Sigma_v$ are real
and momentum independent, and $\Sigma_q$ is negligible. Hence,
\begin{align}
\Sigma_v\equiv\frac{\Sigma_u}{1-\Sigma_q}\sim\Sigma_u,\quad
M^{*}_N\equiv\frac{M_N+\tilde{\Sigma}_s}{1-\Sigma_q}\sim
M_N+\tilde{\Sigma}_s\label{selfe}.
\end{align}

The phenomenological representation of the
nucleon propagator can then be written as
\begin{align}
G(q)=\frac{1}{ \slash
\hspace{-0.2cm}q-M_n-\Sigma(q)}\rightarrow\lambda^2\frac{\slash
\hspace{-0.2cm}q+M^{*}- \slash
\hspace{-0.2cm}u\Sigma_v}{(q_0-E_q)(q_0-\bar{E}_0)}, \label{phen1}
\end{align}
where $\lambda$ is unity in this discussion.  But if one includes
the effect of $\Sigma_q$, $\lambda^2= (1-\Sigma_q)^{-1}$. $E_q$ and
$\bar{E}_q$ are the positive and negative energy poles,
respectively,
\begin{align}
E_q&=\Sigma_v+\sqrt{\vec{q}^2+M_N^{*2}},\label{polep}\\
\bar{E}_q&=\Sigma_v-\sqrt{\vec{q}^2+M_N^{*2}}\label{pole}.
\end{align}
With fixed $\vert \vec{q} \vert$, $G(q)$ depends only on $q_0$. One
can extract self energy near $\sim E_q$ with analytic properties of
the nucleon propagator.

\section{QCD sum rule and matrix elements in the asymmetric nuclear
medium}

\subsection{Operator Product Expansion and Borel sum rule}

To express the self-energies in terms of QCD variables, we start
with analyzing the correlation function via the operator product
expansion (OPE). The Correlator is defined as
\begin{align}
\Pi(q) \equiv i \int d^4 x e^{iqx} \langle\Psi_0\vert
\textrm{T}[\eta(x)\bar{\eta}(0)]\vert\Psi_0\rangle,\label{corr}
\end{align}
where $\eta(x)$ is an interpolating current of the nucleon  and
$\vert\Psi_0\rangle$ is the ground state of the asymmetric nuclear
medium characterized by the rest frame medium density $\rho$, the medium
four-velocity $u^\mu$ and the asymmetry factor $I$. $\vert\Psi_0\rangle$ is assumed to be invariant under parity and time reversal. We
will be using the Ioffe nucleon interpolating current given as in
Ref.~\cite{Ioffe:1981kw,Cohen:1991js},
\begin{align}
\eta(x)=\epsilon_{abc}[u^T_a(x)C\gamma_\mu u_b(x)]\gamma_5\gamma^\mu
d_c(x).\label{crnt}
\end{align}

As in the case of the nucleon propagator, using Lorentz covariance,
parity and time reversal, one can decompose the correlator into
three invariants~\cite{Furnstahl:1992pi}:
\begin{align}
\Pi(q) \equiv \Pi_s(q^2,qu)+\Pi_q(q^2,qu) \slash \hspace{-0.2cm}
q+\Pi_u(q^2,qu) \slash \hspace{-0.2cm}u. \label{corrd}
\end{align}

The three invariants are functions of $q^2$ and $qu$, while the
vacuum invariants depends only on $q^2$. For convenience, we set the
nuclear medium at rest, which means $u^\mu\rightarrow(1,\vec{0})$,
and keep $\vert\vec{q}\vert$ fixed. $\Pi_i(q^2,qu)$ then becomes a
function of $q_0$ only, which means $\Pi_i(q^2,q\cdot
u)\rightarrow\Pi_i(q_0,\vert\vec{q}\vert\rightarrow\textrm{fixed})$
$(i=\{s,q,u\})$.

As mentioned before, we will follow the formalism adopted in
Ref.~\cite{Cohen:1991js} and write the energy dispersion relation
for the invariant functions at fixed three-momentum
$\vert\vec{q}\vert$:
\begin{align}
\Pi_i(q_0,\vert\vec{q}\vert)&=\frac{1}{2\pi i}\int^\infty_{-\infty}
d\omega \frac{\Delta\Pi_i(\omega,\vert\vec{q}\vert)}{\omega-q_0}+\textrm{polynomials},
\label{cosh}\nonumber\\
\\
\Delta\Pi_i(\omega,\vert\vec{q}\vert)&\equiv\lim_{\epsilon\rightarrow0^{+}}
[\Pi_i(\omega+i\epsilon,\vert\vec{q}\vert)-\Pi_i
(\omega-i\epsilon,\vert\vec{q}\vert)]\nonumber\\
&=2\textrm{Im}[\Pi_i(\omega,\vert\vec{q}\vert)]\label{disc}.
\end{align}

The lowest energy contribution to the discontinuity will be
saturated by a quasi-nucleon and quasi-hole contribution in the
positive and negative energy domain respectively. Their contribution
to the spectral density will be given as in Eq.~\eqref{phen1},
which will have the following contribution to the invariant
functions:
\begin{align}
\Pi_s(q_0,\vert\vec{q}\vert)&=-\lambda^{*2}_N\frac{M^{{}*}_N}{(q_0-E_q)(q_0-\bar{E}_q)}+\cdots,\\
\Pi_q(q_0,\vert\vec{q}\vert)&=-\lambda^{*2}_N\frac{1}{(q_0-E_q)(q_0-\bar{E}_q)}+\cdots,\\
\Pi_u(q_0,\vert\vec{q}\vert)&=+\lambda^{*2}_N\frac{\Sigma_v}{(q_0-E_q)(q_0-\bar{E}_q)}+\cdots,
\end{align}
where $\lambda^{*2}_N$ is the residue at the quasi-nucleon pole,
which accounts for the coupling of the interpolating current to the
quasi-nucleon excitation state, and the omitted parts are  the
contributions from the higher excitation states, which will be
accounted for through the continuum contribution after the Borel
transformation.

The even and odd parts of the invariant functions are respectively related to the following parts of the discontinuity:
\begin{align}
\Pi_i^E(q_0^2,\vert\vec{q}\vert)&=\frac{1}{2\pi i}\int^\infty_{-\infty}
d\omega \frac{ \omega \Delta\Pi_i(\omega,\vert\vec{q}\vert)}{\omega^2-q_0^2}+\textrm{polynomials},
\nonumber \\
\Pi_i^O(q_0^2,\vert\vec{q}\vert)&=\frac{1}{2\pi i}\int^\infty_{-\infty}
d\omega \frac{ \Delta\Pi_i(\omega,\vert\vec{q}\vert)}{\omega^2-q_0^2}+\textrm{polynomials}, \label{dis-eo}
\end{align}
where we have defined the invariants with different superscripts from the following decomposition according the the parity in $q_0$:
\begin{align}
\Pi_i(q_0,\vert\vec{q}\vert)=\Pi_i^E(q_0^2,\vert\vec{q}\vert)+q_0
\Pi_i^O(q_0^2,\vert\vec{q}\vert).
\end{align}

The OPE of the three invariants of both the even and odd parts can
be expressed as
\begin{align}
\Pi_i(q^2,q_0^2)=\sum_n C^i_n(q^2,q_0^2)\langle
\hat{O}_n\rangle_{\rho,I},
\end{align}
where $\langle \hat{O}_n\rangle_{\rho,I}$ is the ground state
expectation value of the physical operator in the asymmetric nuclear
medium; $\langle\Psi_0\vert \hat{O}_n\vert\Psi_0\rangle_{\rho,I}$.
We will be adopting the OPE at $q^2 \rightarrow - \infty$ at finite
$\vert\vec{q}\vert\rightarrow\textrm{fixed}$; this is equivalent to
the limit of $q_0^2 \rightarrow - \infty$ at finite
$\vert\vec{q}\vert\rightarrow\textrm{fixed}$. The Wilson
coefficients $C^i_n(q^2,q_0)$ thus can be calculated in QCD at short
time~\cite{Cohen:1991js}.

The OPE of the invariants for the proton interpolating current are
given as follows up to dimension-five operators:
\begin{widetext}\allowdisplaybreaks
\begin{align}
\Pi_s^E(q_0^2,\vert\vec{q}\vert)=&~\frac{1}{4\pi^2}q^2\ln(-q^2)\langle
\bar{d}d \rangle_{\rho,I}+\frac{4}{3\pi^2}\frac{q_0^2}{q^2}\langle
\bar{d} \{iD_{0} iD_{0}\}
d \rangle_{\rho,I}, \\
\Pi_s^O(q_0^2,\vert\vec{q}\vert)=&-\frac{1}{2\pi^2}\ln(-q^2)\langle
\bar{d}iD_0d \rangle_{\rho,I},\\
\Pi_q^E(q_0^2,\vert\vec{q}\vert)=&-\frac{1}{64\pi^4}(q^2)^2\ln(-q^2)+\left[\frac{1}{9\pi^2}\ln(-q^2)-\frac{4}{9\pi^2}\frac{q_0^2}{q^2}\right]\langle
\bar{d} \{\gamma_{0} iD_{0}\} d
\rangle_{\rho,I}+\left[\frac{4}{9\pi^2}\ln(-q^2)-\frac{4}{9\pi^2}\frac{q_0^2}{q^2}\right]\langle
\bar{u}
\{\gamma_{0} iD_{0}\} u \rangle_{\rho,I}\nonumber\\
&-\frac{1}{32\pi^2}\ln(-q^2)\left\langle\frac{\alpha_s}{\pi}G^2\right\rangle_{\rho,I}
-\frac{1}{144\pi^2}\left[\ln(-q^2)-\frac{4q_0^2}{q^2}\right]\left\langle\frac{\alpha_s}{\pi}
[(u\cdot
G)^2+(u\cdot \tilde{G})^2]\right\rangle_{\rho,I}, \\
\Pi_q^O(q_0^2,\vert\vec{q}\vert)=&~
\frac{1}{6\pi^2}\ln(-q^2)\left[\langle u^\dagger
u\rangle_{\rho,I}+\langle d^\dagger
d\rangle_{\rho_I}\right]-\frac{2}{3\pi^2}\frac{q_0^2}{(q^2)^2}\langle
\bar{u} \{\gamma_{0} iD_{0} iD_{0}\} u
\rangle_{\rho,I}-\frac{2}{3\pi^2}\frac{q_0^2}{(q^2)^2}\langle
\bar{d} \{\gamma_{0} iD_{0} iD_{0}\} d
\rangle_{\rho,I}\nonumber\\
&-\frac{2}{3\pi^2}\frac{1}{q^2}\langle \bar{u} \{\gamma_{0} iD_{0}
iD_{0}\} u\rangle_{\rho,I}+\frac{1}{18\pi^2}\frac{1}{q^2}\langle
g_su^\dagger\sigma\cdot
\mathcal{G} u\rangle_{\rho,I},\\
\Pi_u^E(q_0^2,\vert\vec{q}\vert)=&~\frac{1}{12\pi^2}q^2\ln(-q^2)\left[7\langle
u^\dagger u\rangle_{\rho,I}+\langle d^\dagger d
\rangle_{\rho,I}\right]+\frac{3}{\pi^2}\frac{q_0^2}{q^2}\langle
\bar{u} \{\gamma_{0} iD_{0} iD_{0}\} u \rangle_{\rho,I}
+\frac{1}{\pi^2}\frac{q_0^2}{q^2}\langle \bar{d} \{\gamma_{0} iD_{0}
iD_{0}\}
d \rangle_{\rho,I}\nonumber\\
&-\frac{1}{6\pi^2}\ln(-q^2)\langle g_su^\dagger\sigma\cdot
\mathcal{G} u\rangle_{\rho,I}+\frac{1}{12\pi^2}\ln(-q^2)\langle
g_sd^\dagger\sigma\cdot \mathcal{G} d\rangle_{\rho,I}, \\
\Pi_u^O(q_0^2,\vert\vec{q}\vert)=&-\frac{4}{9\pi^2}\ln(-q^2)\langle
\bar{d} \{\gamma_{0} iD_{0}\} d
\rangle_{\rho,I}-\frac{16}{9\pi^2}\ln(-q^2)\langle \bar{u}
\{\gamma_{0} iD_{0}\} u
\rangle_{\rho,I}+\frac{1}{36\pi^2}\ln(-q^2)\left\langle\frac{\alpha_s}{\pi}[(u\cdot
G)^2+(u\cdot \tilde{G})^2]\right\rangle_{\rho,I}.
\end{align}

The quark part and their flavor structure of the above OPE can be
obtained by suitable substitutions of the corresponding OPE for the
$\Sigma$ given in Ref.~\cite{Jin:1994bh}; by changing $q\rightarrow
u$, $s\rightarrow d$, and neglecting terms proportional to $m_s$.
Moreover, when both $u$ and $d$ quarks are identified to the generic
light flavor $q$, our OPE also reduces to that given in
Ref.~\cite{Jin:1993up}.

The next task is to identify the nucleon self-energies in the
asymmetric nuclear medium.  We therefore have to concentrate on the
quasi nucleon pole and not on the quasi hole nor the continuum
excitations. To this end, we apply the Borel transformation with
appropriate weighting function to the dispersion
relation~\cite{Furnstahl:1992pi} and the corresponding differential
operator $\mathcal{B}$ to the OPE side; details of Borel
transformations are given in Appendix E. The Borel transformed
invariants which contain the continuum corrections are as follows:

\begin{align}
\bar{\mathcal{B}}[\Pi_s(q_0^2,\vert\vec{q}\vert)]=&~\lambda_N^{{*}2}M_p^{*}e^{-(E_q^2-\vec{q}^2)/M^2}\nonumber\\
=&-\frac{1}{4\pi^2}(M^2)^2E_1\langle \bar{d}d
\rangle_{\rho,I}-\frac{4}{3\pi^2}\vec{q}^2\langle \bar{d} \{iD_{0}
iD_{0}\} d
\rangle_{\rho,I}L^{-\frac{4}{9}}+\bar{E}_q\bigg[-\frac{1}{2\pi^2}M^2E_0\langle
\bar{d}iD_0d
\rangle_{\rho,I}L^{-\frac{4}{9}}\bigg],\label{borel1} \\
\bar{\mathcal{B}}[\Pi_q(q_0^2,\vert\vec{q}\vert)]=&~ \lambda_N^{{*}2}e^{-(E_q^2-\vec{q}^2)/M^2}\nonumber\\
=&~\frac{1}{32\pi^4}(M^2)^3E_2L^{-\frac{4}{9}}-
\left(\frac{1}{9\pi^2}M^2E_0-\frac{4}{9\pi^2}\vec{q}^2\right)\langle
\bar{d} \{\gamma_{0} iD_{0}\} d
\rangle_{\rho,I}L^{-\frac{4}{9}}\nonumber\\
&-\left(\frac{4}{9\pi^2}M^2E_0-\frac{4}{9\pi^2}\vec{q}^2\right)\langle
\bar{u} \{\gamma_{0} iD_{0}\} u
\rangle_{\rho,I}L^{-\frac{4}{9}}+\frac{1}{32\pi^2}M^2\left\langle\frac{\alpha_s}{\pi}G^2 \right\rangle_{\rho,I}E_0L^{-\frac{4}{9}} \nonumber\\
&+\frac{1}{144\pi^2}\left(M^2E_0
-4\vec{q}^2\right)\left\langle\frac{\alpha_s}{\pi}[(u\cdot
G)^2+(u\cdot
\tilde{G})^2]\right\rangle_{\rho,I}L^{-\frac{4}{9}}+\bar{E}_q\bigg[\frac{1}{6\pi^2}M^2E_0L^{-\frac{4}{9}}\left[\langle
u^\dagger u\rangle_{\rho,I}+\langle d^\dagger d
\rangle_{\rho,I}\right]\nonumber\\
&-\frac{2}{3\pi^2} \left(1-\frac{\vec{q}^2}{M^2}\right)\langle
\bar{u} \{\gamma_{0} iD_{0} iD_{0}\} u
\rangle_{\rho,I}L^{-\frac{4}{9}}-\frac{2}{3\pi^2}\left(1-\frac{\vec{q}^2}{M^2}\right)\langle
\bar{d} \{\gamma_{0} iD_{0} iD_{0}\} d
\rangle_{\rho,I}L^{-\frac{4}{9}}\nonumber\\
&-\frac{2}{3\pi^2}\langle \bar{u} \{\gamma_{0} iD_{0} iD_{0}\} u
\rangle_{\rho,I}L^{-\frac{4}{9}}+\frac{1}{18\pi^2}\langle
g_su^\dagger\sigma\cdot \mathcal{G}
u\rangle_{\rho,I}L^{-\frac{4}{9}}\bigg],\label{borel2} \\
\bar{\mathcal{B}}[\Pi_u(q_0^2,\vert\vec{q}\vert)]=&~\lambda_N^{{*}2}\Sigma_v^p e^{-(E_q^2-\vec{q}^2)/M^2}\nonumber\\
=&~\frac{1}{12\pi^2}(M^2)^2\left[7\langle u^\dagger
u\rangle_{\rho,I}+\langle d^\dagger d
\rangle_{\rho,I}\right]E_1L^{-\frac{4}{9}}+\frac{3}{\pi^2}\vec{q}^2\langle
\bar{u} \{\gamma_{0} iD_{0} iD_{0}\} u
\rangle_{\rho,I}L^{-\frac{4}{9}}\nonumber\\
&+\frac{1}{\pi^2}\vec{q}^2\langle \bar{d} \{\gamma_{0} iD_{0}
iD_{0}\} d
\rangle_{\rho,I}L^{-\frac{4}{9}}-\frac{1}{6\pi^2}M^2\langle
g_su^\dagger\sigma\cdot \mathcal{G}
u\rangle_{\rho,I}E_0L^{-\frac{4}{9}}+\frac{1}{12\pi^2}M^2\langle
g_sd^\dagger\sigma\cdot \mathcal{G} d\rangle_{\rho,I}E_0L^{-\frac{4}{9}}\nonumber\\
&+\bar{E}_q\bigg[\frac{4}{9\pi^2}M^2\langle \bar{d} \{\gamma_{0}
iD_{0}\} d
\rangle_{\rho,I}E_0L^{-\frac{4}{9}}+\frac{16}{9\pi^2}M^2\langle
\bar{u} \{\gamma_{0} iD_{0}\} u
\rangle_{\rho,I}E_0L^{-\frac{4}{9}}\nonumber\\
& -\frac{1}{36\pi^2}M^2\left\langle\frac{\alpha_s}{\pi}[(u\cdot
G)^2+(u\cdot
\tilde{G})^2]\right\rangle_{\rho,I}E_0L^{-\frac{4}{9}}\bigg].
\label{borel3}
\end{align}
\end{widetext}
Here, we include the corrections from the anomalous dimensions as
\begin{align}
L^{-2\Gamma_\eta+\Gamma_{O_n}}\equiv\left[\frac{\ln(M/\Lambda_{QCD})}{\ln(\mu/\Lambda_{QCD})}\right]^{-2\Gamma_\eta+\Gamma_{O_n}},
\end{align}
where $\Gamma_\eta$ ($\Gamma_{O_n}$) is the anomalous dimension of
the interpolating  current $\eta$  ($\hat{O}_n$),  $\mu$ is the
normalization point of the OPE, and $\Lambda_{QCD}$ is the QCD
scale~\cite{Furnstahl:1992pi,Jin:1993up}.

Also, the continuum corrections are taken into account through the
factors
\begin{align}
E_0&\equiv1-e^{s_0^{*}/M^2},\\
E_1&\equiv1-e^{s_0^{*}/M^2}\left(s_0^{*}/M^2+1\right),\\
E_2&\equiv1-e^{s_0^{*}/M^2}\left(s_0^{*2}/2M^4+s_0^{*}/M^2+1\right),
\end{align}
where $s_0^{*}\equiv\omega_0^{2}-\vec{q}^2$, and $\omega_0$ is the
energy at the continuum threshold.
We choose the continuum to be the same as the vacuum value $\omega_0=1.5~\textrm{GeV}$.  This assumption will be justified later as the
results do not have strong $\omega_0$ dependence.

\subsection{Condensates in the asymmetric nuclear medium}

To estimate the matrix elements, we will use the linear density
approximation in the asymmetric nuclear matter,
\begin{align}
\langle\hat{O}\rangle_{\rho,I}=&~\langle\hat{O}\rangle_{\textrm{vac}}+\langle
n\vert\hat{O}\vert n\rangle \rho_n+\langle p \vert \hat{O} \vert p
\rangle\rho_p \nonumber \\
=&~\langle\hat{O}\rangle_{\textrm{vac}}+\frac{1}{2}  (\langle
n\vert\hat{O}\vert n\rangle +\langle p \vert \hat{O} \vert p \rangle
)\rho   \nonumber \\
&+\frac{1}{2} (\langle n\vert\hat{O}\vert n\rangle -\langle p \vert
\hat{O} \vert p \rangle) I \rho . \label{condensatea}
\end{align}

The quark flavor of condensate becomes important in the asymmetric
nuclear medium. Consider an operator $\hat{O}_{u,d}$ composed of
either $u$ or $d$ quarks, respectively.  Making use of the isospin
symmetry relation,
\begin{align}
\langle n\vert\hat{O}_{u,d}\vert n\rangle=\langle
p\vert\hat{O}_{d,u}\vert p\rangle,
\end{align}
we can convert the neutron expectation value to the proton
expectation value, thereby rewriting Eq.~\eqref{condensatea} for the
two-quark operators as follows:
\begin{align}
\langle\hat{O}_{u,d}\rangle_{\rho,I}
=&\langle\hat{O}_{u,d}\rangle_{\textrm{vac}}+(\langle
p\vert\hat{O}_0\vert p\rangle\mp\langle p \vert \hat{O}_1 \vert p
\rangle I)\rho.\label{condensateb}
\end{align}
Here, ``$-$"  and  ``$+$" are for the $u$ and $d$ quark flavors,
respectively, and the isospin operators are defined as
\begin{align}
\hat{O}_0\equiv\frac{1}{2}(\hat{O}_u+\hat{O}_d),\quad
\hat{O}_1\equiv\frac{1}{2}(\hat{O}_u-\hat{O}_d).
\end{align}
Hence,  we will convert all the expectation values in terms of the proton counterparts and  denote them as $\langle p \vert \hat{O}
\vert p \rangle\rightarrow\langle\hat{O}\rangle_p$, throughout this
paper. The next task is to find $\langle\hat{O}_0\rangle_p$ and
$\langle\hat{O}_1\rangle_p$ for all operators appearing in our OPE.

\subsubsection{$\langle \bar{q} D_{\mu_1} \cdots D_{\mu_n} q \rangle$ type of condensates}

Let us start by estimating the lowest-dimensional operators
$\langle[\bar{q}q]_0\rangle_p$ and $\langle[\bar{q}q]_1\rangle_p$.
To find $\langle[\bar{q}q]_1\rangle_p$, we will use an estimate
based on using the QCD energy momentum tensor in the baryon octet
mass relation to leading order in the quark mass \cite{Choi:1991};
Eq.~\eqref{baryonoctet} in Appendix A.  Using Eq.~\eqref{ratio1},
one finds
\begin{align}
\langle[\bar{q}q]_1\rangle_p=&~\frac{1}{2}(\langle
p\vert\bar{u}u\vert p\rangle-\langle p\vert\bar{d}d\vert
p\rangle)\nonumber\\
=&~\frac{1}{2}\left[\frac{(m_{\Xi^{0}}+m_{\Xi^{-}})
-(m_{\Sigma^{+}}+m_{\Sigma^{-}})}{2m_s-2m_q}\right].\label{diff}
\end{align}
We will use the baryon masses as given in the Particle Data
Group~\cite{Nakamura:2010zzi}:
$m_{\Xi^{0}}=1315\textrm{\phantom{1}MeV}$,
$m_{\Xi^{-}}=1321\textrm{\phantom{1}MeV}$,
$m_{\Sigma^{+}}=1190\textrm{\phantom{1}MeV}$,
$m_{\Sigma^{-}}=1197\textrm{\phantom{1}MeV}$.  Using
$m_s=150\textrm{\phantom{1}MeV}$ and
$m_q\equiv\frac{1}{2}(m_u+m_d)=$ 5 MeV, Eq.~\eqref{diff} becomes
\begin{align}
\langle[\bar{q}q]_1\rangle_p=\frac{1}{2}\left(\frac{249~\textrm{MeV}}
{300~\textrm{MeV}-2m_q}\right) \sim 0.43.
\end{align}

For $\langle[\bar{q}q]_0\rangle_p$, we make use of the nucleon
$\sigma_N=45$ MeV term,
\begin{align}
\langle[\bar{q}q]_0\rangle_p=\frac{1}{2}\left(\langle
p\vert\bar{u}u\vert p\rangle+\langle p\vert\bar{d}d\vert
p\rangle\right)=\frac{\sigma_N}{2m_q}\sim 4.5.
\end{align}

For convenience, one can introduce the parameter $\mathcal{R}_{\pm}(m_q)$ defined as
\begin{align}
\langle p\vert\bar{u} u\vert p\rangle\pm\langle p\vert\bar{d} d\vert
p\rangle=&~\mathcal{R}_{\pm}(m_q)\langle p\vert\bar{u} u\vert
p\rangle,
\end{align}
which leads to
\begin{align}
\langle[\bar{q}q]_1\rangle_p=&~\frac{\mathcal{R}_{-}(m_q)}{\mathcal{R}_{+}(m_q)}
\langle[\bar{q}q]_0\rangle_p.\label{asymmcondensate}
\end{align}
Using the previously selected values with the explicit quark mass dependence, we have
\begin{align}
\mathcal{R}_{\pm}(m_q)\equiv&\bigg[1\pm\left(\frac{\sigma_N}{m_q}
-\frac{249~\textrm{MeV}}{300~\textrm{MeV}-2m_q}\right)\bigg/
\nonumber\\
&\left(\frac{\sigma_N}{m_q}
+\frac{249~\textrm{MeV}}{300~\textrm{MeV}-2m_q}\right)\bigg],
\label{def-r-}
\end{align}
so $\mathcal{R}_{\pm}(m_q=5 ~{\rm MeV}))=1\pm0.68$.

Using this parametrization, we can express the $u$ quark or $d$
quark condensates as follows,
\begin{align}
\langle[\bar{q}q]_{u,d}\rangle_{\rho,I}=
&~\langle[\bar{q}q]_{u,d}\rangle_{\textrm{vac}}+\left[1\mp
\frac{\mathcal{R}_{-}(m_q)}{\mathcal{R}_{+}(m_q)}
I\right]\langle[\bar{q}q]_0 \rangle_p\rho,
\end{align}
where $[\bar{q}q]_{u}=\bar{u}u$ and $[\bar{q}q]_{d}=\bar{d}d$. For
$\langle\bar{q}q\rangle_{\textrm{vac}}$, we use the
Gellmann-Oakes-Renner relation:
\begin{align}
2m_q\langle\bar{q}q\rangle_{\textrm{vac}}=-m^2_\pi f^2_\pi,
\end{align}
where $m_\pi=138$ $\textrm{MeV}$ and $f_\pi=98$ $\textrm{MeV}$
\cite{Jin:1993up}. For $m_q=5$ ${\rm MeV}$, we have
$\langle\bar{q}q\rangle_{\textrm{vac}}=-(263~{\rm MeV})^3$.

Likewise, we will further assume that the ratios between the isospin
singlet and triplet operators remain the same for all two quark operator expectation
values with any number of covariant derivatives inserted:
\begin{align}
\langle[\bar{q} D_{\mu_1} \cdots D_{\mu_n}
q]_1\rangle_p=\frac{\mathcal{R}_{-}(m_q)}{\mathcal{R}_{+}(m_q)}
\langle[\bar{q} D_{\mu_1} \cdots D_{\mu_n} q]_0\rangle_p.
\end{align}
With this assumption, $\langle \bar{q} D_{\mu_1} \cdots D_{\mu_n}q
\rangle_{\rho,I}$ can be written as
\begin{align}
\langle[\bar{q}& D_{\mu_1} \cdots D_{\mu_n}
q]_{u,d}\rangle_{\rho,I}\nonumber\\
=& ~\langle[\bar{q} D_{\mu_1} \cdots D_{\mu_n}
q]_{u,d}\rangle_{\textrm{vac}}\nonumber\\
&+\left(1\mp \frac{\mathcal{R}_{-}(m_q)}{\mathcal{R}_{+}(m_q)}
I\right)\langle[\bar{q} D_{\mu_1} \cdots D_{\mu_n} q]_0
\rangle_p\rho.
\end{align}

The symmetric and traceless part of the above type of expectation
values, constitute the moments of the twist-3 $e_n(x,\mu^2)$
structure function  defined as follows~\cite{Jaffe:1991ra}:
\begin{align}
\langle[\bar{q} \{ D_{\mu_1} \cdots D_{\mu_n} \}q]_0\rangle_p
&\equiv(-i)^n
e_n(\mu^2)\{p_{\mu_1}\cdots p_{\mu_n}\},\\
e_n(\mu^2)&\equiv\int dx~ x^n e_n(x,\mu^2),
\end{align}
where $\{\mu_1\cdots\mu_n\}$ means symmetric and traceless indices.
The two-quark twist-3 condensates in our sum rule then can be
written as follows:
\begin{align}
\langle[\bar{q}iD_{\mu'} q]_{u,d}\rangle_{\rho,I} = &~\langle[\bar{q}iD_0q]_{u,d}\rangle_{\rho,I} u'_\mu\nonumber\\
=&~ m_q\langle[q^\dagger q]_{u,d}\rangle_{\rho,I}=0,\\
\langle [\bar{q} \{iD_{\mu'} iD_{\nu'}\} q]_{u,d}
\rangle_{\rho,I}=&~\frac{4}{3}\langle [\bar{q} \{iD_{0} iD_{0}\}
q]_{u,d} \rangle_{\rho,I}\nonumber\\
&\times\left(u'_\mu u'_\nu-\frac{1}{4}g_{\mu\nu}\right),
\end{align}
where the in-medium rest frame $u'_\mu\equiv(1,\vec{0})$ has been taken
and the matrix element is estimated as
\begin{align}
\langle [\bar{q}\{ iD_{\mu'} iD_{\nu'}\} q]_{0}
\rangle_p=~M_N^2e_2(\mu^2)\left(u'_\mu
u'_\nu-\frac{1}{4}g_{\mu\nu}\right),
\end{align}
where one can identify that $M_N^2e_2(\mu^2)=\frac{4}{3}\langle
[\bar{q} \{iD_{0} iD_{0}\} q]_{0} \rangle_{p}$, and $\langle [\bar{q}
\{iD_{0} iD_{0}\} q]_{u,d} \rangle_{\rho,I}$ can be written as
\begin{align}
\langle [\bar{q} \{iD_{0} iD_{0}\} q]_{u,d}
\rangle_{\rho,I}\simeq\left[1\mp\frac{\mathcal{R}_{-}(m_q)}{\mathcal{R}_{+}(m_q)}
I\right] M_N^2e_2(\mu^2)\rho.
\end{align}
Since there are no measurements on the twist-3 structure function,
we will take the estimate for  $M_N^2e_2(\mu^2) \sim
0.3~\textrm{GeV}^2$ given in Ref.~\cite{Jin:1993up,Belyaev:1982sa}.

When spin indices are contracted, the operator becomes
\begin{align}
\langle [\bar{q} D^2 q]_{u,d} \rangle_{\rho,I}=&~\frac{1}{2}\langle
[g_s\bar{q}\sigma \cdot\mathcal{G}q]_{u,d}
\rangle_{\rho,I}\nonumber\\
=&~\frac{1}{2}
\left[1\mp\frac{\mathcal{R}_{-}(m_q)}{\mathcal{R}_{+}(m_q)}
I\right]\langle[g_s\bar{q}\sigma\cdot\mathcal{G}q]_0\rangle_p\rho,
\end{align}
where $\langle[g_s\bar{q}\sigma\cdot\mathcal{G}q]_0\rangle_{p}$ is
chosen to be $3~\textrm{GeV}^2$ as in
Ref.~\cite{Jin:1993up,Belyaev:1982sa}.

\subsubsection{$\langle \bar{q} \gamma_{\mu_1} D_{\mu_2} \cdots D_{\mu_n} q \rangle$ type of condensates}

The simplest condensate of this type is
\begin{align}
\langle \bar{q} \gamma_\lambda q \rangle_{\rho,I}=\langle \bar{q}
\slash \hspace{-0.2cm}  u' q \rangle_{\rho,I}
u'_\lambda\rightarrow\langle q^\dagger q \rangle_{\rho,I}u'_\lambda.
\end{align}
For this, the ratio $\langle u^\dagger u\rangle_p / \langle d^\dagger
d\rangle_p=2$, and the isospin relation for $\langle
q^\dagger q\rangle_{\rho,I}$ can be written as
\begin{align}
\langle[q^\dagger q]_1\rangle_p =\frac{1}{3}\langle[q^\dagger
q]_0\rangle_p,
\end{align}
which leads to the following matrix elements appearing in the sum
rule:
\begin{align}
\langle [q^\dagger q]_{u,d}\rangle_{\rho,I}=&\left(1\mp
\frac{1}{3}I\right)\langle  [q^\dagger q]_0
\rangle_p\rho=\left(\frac{3}{2}\mp\frac{1}{2}I\right)\rho.
\end{align}

When covariant derivatives are included, one can estimate the
two-quark twist-2 condensates from the corresponding parton
distribution function,
\begin{align}
\langle \bar{q} \{ \gamma_{\mu_1} D_{\mu_2} \cdots D_{\mu_n}\} q
\rangle_p\equiv\frac{(-i)^{n-1}}{2M_N}A^q_n(\mu^2)\{p_{\mu_1}\cdots
p_{\mu_n}\},
\end{align}
where $A^q_n(\mu^2)=[A^u_n(\mu^2)+A^d_n(\mu^2)]/2$ is the reduced
matrix element~\cite{Collins:1981uw,Curci:1980uw}:
\begin{align}
A^q_n(\mu^2)=2\int^1_0
dx~x^{n-1}[q(x,\mu^2)+(-1)^n\bar{q}(x,\mu^2)],
\end{align}
where $q(x,\mu^2)$ and $\bar{q}(x,\mu^2)$ are the distribution
functions for quarks and antiquarks in the proton, respectively, and
$\mu^2$ is the renormalization scale. For the distribution
functions, we used the leading order (LO) parametrization given in
Ref.~\cite{Gluck:1991ng}.

Specifically, the spin-2 part can be written
as~\cite{Hatsuda:1991ez}
\begin{align}
\langle [\bar{q} \{\gamma_\mu& iD_\nu\} q]_{u,d}
\rangle_{\rho,I}\nonumber\\
\rightarrow &~\langle [\bar{q}
\{\gamma_{\mu'}
iD_{\nu'}\} q]_{u,d} \rangle_{\rho,I}\nonumber\\
=&~\frac{4}{3}\langle [\bar{q} \{\gamma_{0} iD_{0}\} q]_{u,d}
\rangle_{\rho,I}\left(u'_\mu u'_\nu-\frac{1}{4}g_{\mu\nu}\right),
\end{align}
where the in-medium rest frame has been taken. The matrix elements
for each flavor in $\langle [\bar{q} \{\gamma_{0} iD_{0}\} q]_{u,d}
\rangle_{p}$ can be identified as
\begin{align}
\langle \bar{u} \{\gamma_{\mu'} iD_{\nu'}\} u
\rangle_p=\frac{1}{2}M_N A^u_2(\mu^2)\left(u'_\mu
u'_\nu-\frac{1}{4}g_{\mu\nu}\right),\\
\langle \bar{d} \{\gamma_{\mu'} iD_{\nu'}\} d
\rangle_p=\frac{1}{2}M_N A^d_2(\mu^2)\left(u'_\mu
u'_\nu-\frac{1}{4}g_{\mu\nu}\right),
\end{align}
where $A^u_2(\mu^2)\simeq0.74$ and $A^d_2(\mu^2)\simeq0.36$ at
$\mu^2=0.25$ $\textrm{GeV}^2$ (LO)~\cite{Gluck:1991ng}.

One can introduce a ratio factor for  $\langle\hat{O}_1\rangle_p$ as
\begin{align}
\langle [\bar{q} \{\gamma_{\mu'} iD_{\nu'}\} q]_{1}
\rangle_{p}=~\mathcal{R}_{A_2}(\mu^2)\langle [\bar{q}
\{\gamma_{\mu'} iD_{\nu'}\} q]_{0} \rangle_{p},
\end{align}
where
$\mathcal{R}_{A_2}(\mu^2)=(A^u_2-A^d_2)/(A^u_2+A^d_2)\simeq0.35$ so
$\langle [\bar{q} \{\gamma_{0} iD_{0}\} q]_{u,d} \rangle_{\rho,I}$
can be written as
\begin{align}
\langle [\bar{q} \{\gamma_{0} iD_{0}\} q]_{u,d}
\rangle_{\rho,I}=&~\left[1\mp \mathcal{R}_{A_2}(\mu^2)
I\right]\langle
[\bar{q} \{\gamma_{0} iD_{0}\} q]_{0} \rangle_{p}\rho\nonumber\\
=&~\left[1\mp \mathcal{R}_{A_2}(\mu^2) I\right]\frac{1}{2}M_N
A^q_2(\mu^2) \rho.
\end{align}

The spin-3 part can be written as
\begin{align}
\langle [\bar{q} \{\gamma_{\lambda'} &iD_{\mu'} iD_{\nu'}\} q]_{u,d}
\rangle_{\rho,I}\nonumber\\
=&~2\langle [\bar{q} \{\gamma_{0} iD_{0} iD_{0}\} q]_{u,d}
\rangle_{\rho,I}\nonumber\\
&\times\bigg[u'_\lambda u'_\mu u'_\nu-\frac{1}{6}(u'_\lambda
g_{\mu\nu}+u'_\mu g_{\lambda\nu}+u'_\nu g_{\lambda\mu})\bigg],
\end{align}
where the matrix elements for each flavor in $\langle [\bar{q}
\{\gamma_{0} iD_{0} iD_{0} \} q]_{u,d} \rangle_{p}$ can be
identified with
\begin{align}
\langle \bar{u} \{&\gamma_{\lambda'} iD_{\mu'} iD_{\nu'}\} u
\rangle_p\nonumber\\
=&~\frac{1}{2}M_N^2 A^u_3(\mu^2)\nonumber\\
&\times\bigg[u_{\lambda'} u_{\mu'} u_{\nu'}-\frac{1}{6}(u_{\lambda'}
g_{\mu'\nu'}+u_{\mu'}
g_{\lambda'\nu'}+u_{\nu'} g_{\lambda'\mu'})\bigg],\\
\langle \bar{d} \{&\gamma_{\lambda'} iD_{\mu'} iD_{\nu'}\} d
\rangle_p\nonumber\\
=&~\frac{1}{2}M_N^2 A^d_3(\mu^2)\nonumber\\
&\times\bigg[u_{\lambda'} u_{\mu'} u_{\nu'}-\frac{1}{6}(u_{\lambda'}
g_{\mu'\nu'}+u_{\mu'} g_{\lambda'\nu'}+u_{\nu'}
g_{\lambda'\mu'})\bigg],
\end{align}
where $A^u_3(\mu^2)\simeq0.22$ and $A^d_3(\mu^2)\simeq0.07$ at
$\mu^2=0.25$ $\textrm{GeV}^2$ (LO)~\cite{Gluck:1991ng}. Similar to
the spin-2 condensate case, one can write
$\langle\hat{O}_1\rangle_p$ for spin-3 condensate as
\begin{align}
\langle [\bar{q} \{\gamma_{\lambda'} iD_{\mu'} iD_{\nu'}\} q]_{1}
\rangle_{p}=\mathcal{R}_{A_3}(\mu^2)\langle [\bar{q}
\{\gamma_{\lambda'} iD_{\mu'} iD_{\nu'}\} q]_{0} \rangle_{p},
\end{align}
where
$\mathcal{R}_{A_3}(\mu^2)=(A^u_3-A^d_3)/(A^u_3+A^d_3)\simeq0.51$, and
$\langle [\bar{q} \{\gamma_{0} iD_{0} iD_{0}\} q]_{u,d}
\rangle_{\rho,I}$ can be written as
\begin{align}
\langle [\bar{q} \{&\gamma_{0} iD_{0} iD_{0} \} q]_{u,d}
\rangle_{\rho,I}\nonumber\\
=&~\left[1\mp \mathcal{R}_{A_3}(\mu^2)
I\right]\langle
[\bar{q} \{\gamma_{0} iD_{0} iD_{0}\} q]_{0} \rangle_{p}\rho\nonumber\\
=&~\left[1\mp \mathcal{R}_{A_3}(\mu^2) I\right]]\frac{1}{2}M_N^2
A^q_3(\mu^2) \rho.
\end{align}

Operators with contracted spin indices are
\begin{align}
\langle [\bar{q} \slash \hspace{-0.2cm}  D q]_{u,d} \rangle_{\rho,I}=&~0,\\
\langle [q^\dagger  D^2 q]_{u,d}
\rangle_{\rho,I}=&~\frac{1}{2}\langle[g_sq^\dagger\sigma\cdot\mathcal{G}q]_{u,d}\rangle_{\rho,I}\nonumber\\
\simeq&~\frac{1}{2}\left(1\mp \mathcal{R}_{A_3}I\right)\langle [g_s
q^\dagger\sigma\cdot\mathcal{G}q]_0 \rangle_p\rho,
\end{align}
where $\langle[g_s q^\dagger\sigma\cdot\mathcal{G}q]_0 \rangle_p$ is
chosen to be $-0.33$
$\textrm{GeV}^2$~\cite{Jin:1993up,Belyaev:1982sa}.

\subsubsection{Gluon condensates}
As for the gluon operators, because they do not carry quark flavors,
the expectation values do not depend on $I$. These operators can be
written as~\cite{Jin:1992id,Jin:1993up}
\begin{align}
&\left\langle\frac{\alpha_s}{\pi}G^2\right\rangle_{\rho,I}= \left\langle\frac{\alpha_s}{\pi}G^2\right\rangle_{\textrm{vac}}-2\left\langle\frac{\alpha_s}{\pi}(\vec{E}^2-\vec{B}^2)\right\rangle_{p}\rho,\\
&\left\langle\frac{\alpha_s}{\pi}[(u\cdot G)^2+(u\cdot
\tilde{G})^2]\right\rangle_{\rho,I}=-
\left\langle\frac{\alpha_s}{\pi}(\vec{E}^2+\vec{B}^2)\right\rangle_{p}\rho,
\end{align}
where $\vec{E}$ and $\vec{B}$ are the color electric and color
magnetic fields. For the expectation values of the gluon operators
we take;
$\langle(\alpha_s/\pi)G^2\rangle_{\textrm{vac}}=(0.33~\textrm{GeV})^4$
\cite{Shifman:1978bx},
$\langle(\alpha_s/\pi)(\vec{E}^2-\vec{B}^2)\rangle_{p}=0.325\pm0.075~\textrm{GeV}$,
and $\langle(\alpha_s/\pi)(\vec{E}^2+\vec{B}^2)\rangle_{p}=0.10
\pm0.01~\textrm{GeV}$~\cite{Jin:1992id}.

\subsection{Dimension 6 four-quark operators}

In many previous QCD sum rule studies, dimension-six four-quark
condensates are assumed to have the factorized form as
\begin{align}
\left\langle u^a_\alpha\bar{u}^b_\beta
u^c_\gamma\bar{u}^d_\delta\right\rangle_{\rho,I}\simeq&\left\langle
u^a_\alpha\bar{u}^b_\beta\right\rangle_{\rho,I}\left\langle
u^c_\gamma\bar{u}^d_\delta\right\rangle_{\rho,I}\nonumber\\
&-\left\langle
u^a_\alpha\bar{u}^d_\delta\right\rangle_{\rho,I}\left\langle
u^c_\gamma\bar{u}^b_\beta\right\rangle_{\rho,I},\\
\left\langle u^a_\alpha\bar{u}^b_\beta
d^c_\gamma\bar{d}^d_\delta\right\rangle_{\rho,I}\simeq&\left\langle
u^a_\alpha\bar{u}^b_\beta\right\rangle_{\rho,I} \left\langle
d^c_\gamma\bar{d}^d_\delta\right\rangle_{\rho,I}.\label{factorization}
\end{align}
While large $N_c$ arguments can be made to justify factorization in
the vacuum, no such argument exists in the medium. For the in-medium
case, a renewed approach was developed in which the in-medium
four-quark condensates are evaluated within the PCQM
\cite{Drukarev:2003xd,Drukarev:2004fn,Drukarev:2012av}. In this
method, the vacuum condensates are factorized as in
Eq.~\eqref{factorization} but in-medium terms are evaluated by
including intermediate states that include pion clouds. There are
some previous  results to calculate the four-quark operators
appearing in the nucleon OPE.  For example, in Ref.
~\cite{Drukarev:2012av}, the expectation values were calculated
within the PCQM.  Another approach uses a Fierz rearrangement
suitable for factorization as in our case~\cite{Thomas:2007gx}.

In this study, after using the Fierz transformation as above, for
the scalar four-quark operators, we change the four-quark operators
to vary from a mild factorized form to a density independent limit
that preserves the consistent nucleon sum rule as in
Ref.~\cite{Cohen:1994wm}. For the spin-2 four-quark (twist-4)
operators, we use a Fierz rearrangement to extract the independent
four-quark operators that can be related to higher twist effects in
DIS data. Using the following steps, we have classified the
four-quark condensates in terms of the independent operators and of
different twist.

\subsubsection{Twist-4 operators with a single quark flavor}
The first type of four-quark operator appearing in the OPE of the
nucleon sum rule involves quark operators with the same flavor and
is of the color anti triplet diquark times triplet anti diquark
form. Using the following Fierz transformation, one can identify the
independent four-quark operators in terms of products of
quark-antiquark pairs,

\begin{align}
 \epsilon_{abc}&\epsilon_{a'b'c}(u^T_a C\gamma_\mu
u_b)(\bar{u}_{b'} \gamma_\nu C \bar{u}^T_{a'})\nonumber\\
=&~\epsilon_{abc}\epsilon_{a'b'c}~\frac{1}{16}(\bar{u}_{a'} \Gamma^o
u_{a})(\bar{u}_{b'} \Gamma^k u_{b})\textrm{Tr}\left[
\gamma_\mu \Gamma_k \gamma_\nu C \Gamma_o^T C\right]\nonumber\\
=&~\epsilon_{abc}\epsilon_{a'b'c}~\frac{1}{16}~\bigg\{(\bar{u}_{a'}u_a)(\bar{u}_{b'}u_b)(-4g_{\mu\nu})\nonumber\\
&+
(\bar{u}_{a'}\gamma_5u_a)(\bar{u}_{b'}\gamma_5u_b)(4g_{\mu\nu})\nonumber\\
& +(\bar{u}_{a'}\gamma^{\alpha}u_a)(\bar{u}_{b'}\gamma^{\beta}
u_b) (4S_{\mu\beta\nu\alpha})\nonumber\\
&-(\bar{u}_{a'}\gamma^{\alpha}\gamma_5u_a)(\bar{u}_{b'}\gamma^{\beta} \gamma_5 u_b)(4S_{\mu\beta\nu\alpha})\nonumber\\
&+(\bar{u}_{a'}\sigma^{\alpha\bar{\alpha}}u_a)
(\bar{u}_{b'}\sigma^{\beta\bar{\beta}}u_b)\frac{1}{4}\textrm{Tr}\left[
\gamma_\mu \sigma_{\beta\bar{\beta}} \gamma_\nu
\sigma_{\alpha\bar{\alpha}}\right]\nonumber\\
&+(\bar{u}_{a'}\gamma^{\alpha}u_a)
(\bar{u}_{b'}\gamma^{\beta}\gamma_5
u_b)(8i\epsilon_{\mu\beta\nu\alpha})\nonumber\\
&-(\bar{u}_{a'}u_a) (\bar{u}_{b'}\sigma^{\alpha\bar{\alpha}}u_b)(8i
g_{\alpha\mu}g_{\bar{\alpha}\nu})\nonumber\\
&-(\bar{u}_{a'}\gamma_5u_a)
(\bar{u}_{b'}\sigma^{\alpha\bar{\alpha}}u_b)
(4\epsilon_{\mu\nu\alpha\bar{\alpha}})\bigg\}, \label{ope-4q-1}
\end{align}
where $\Gamma=\{I,\gamma_\alpha,i\gamma_\alpha\gamma_5,
\sigma_{\alpha\beta},\gamma_5\}$ and
$S_{\mu\alpha\nu\beta}=g_{\mu\alpha}g_{\nu\beta}+
g_{\mu\beta}g_{\alpha\nu}-g_{\mu\nu}g_{\alpha\beta}$.

When quarks of the same flavor combine into a diquark, certain
combinations are not allowed due to Fermi statistics. From these
conditions, one can extract constraints among four-quark operators
that can be used to identify independent operators. Among several
conditions, the most suitable constraint for our OPE can be obtained
from the zero identity used in Ref.~\cite{Thomas:2007gx}. With the
constraint Eq.~\eqref{constraint1} in Appendix B,
Eq.~\eqref{ope-4q-1} can be simplified as
\begin{align}
\epsilon_{abc}&\epsilon_{a'b'c}(u^T_a C\gamma_\mu
u_b)(\bar{u}_{b'} \gamma_\nu C \bar{u}^T_{a'})\nonumber\\
=&~\epsilon_{abc}\epsilon_{a'b'c}\frac{1}{16}\bigg\{[(\bar{u}_{a'}\gamma^{\alpha}u_a)(\bar{u}_{b'}\gamma^{\beta}
u_b)\nonumber\\
&-(\bar{u}_{a'}\gamma^{\alpha}\gamma_5u_a)(\bar{u}_{b'}\gamma^{\beta} \gamma_5 u_b)](8S_{\mu\beta\nu\alpha})\nonumber\\
&+(\bar{u}_{a'}\gamma^{\alpha}u_a)(\bar{u}_{b'}\gamma^{\beta}\gamma_5
u_b)(16 i\epsilon_{\mu\beta\nu\alpha})\bigg\}.\label{4qope1}
\end{align}
The last term in Eq.~\eqref{4qope1} will be dropped as one should
take an expectation value with respect to a parity-even nuclear
medium ground state. Then only two types of four-quark operators
remain. Each type can be written as
\begin{align}
&\epsilon_{abc}\epsilon_{a'b'c}(\bar{q}_{a'} \Gamma^\alpha
q_a)(\bar{q}_{b'} \Gamma^{\beta}
q_b)\nonumber\\
&=\epsilon_{abc}\epsilon_{a'b'c}\bigg\{\frac{1}{9}\delta_{a'a}\delta_{b'b}(\bar{q}
\Gamma^\alpha q)(\bar{q} \Gamma^{\beta}
q)\nonumber\\
&~~+\frac{2}{3}t^A_{aa'}\delta_{b'b}(\bar{q} \Gamma^\alpha t^A
q)(\bar{q} \Gamma^{\beta}
q)+\frac{2}{3}\delta_{a'a}t^B_{bb'}(\bar{q} \Gamma^\alpha q)(\bar{q}
\Gamma^{\beta} t^B q)\nonumber\\
&~~+4t^A_{aa'} t^B_{bb'}(\bar{q} \Gamma^\alpha t^A q)(\bar{q}
\Gamma^{\beta} t^B q)\bigg\},\label{color-decomposition1}
\end{align}
where $\Gamma^\alpha=\{\gamma^\alpha,i\gamma^\alpha\gamma_5\}$ and
$t^A$ is the generator of SU(3) normalized as
$\textrm{Tr}[t^At^B]=\frac{1}{2}\delta^{AB}$. Combined with the
product of epsilon tensors
$\epsilon_{abc}\epsilon_{a'b'c}=\delta_{bb'}\delta_{aa'}-\delta_{ba'}\delta_{ab'}$,
one finds that the second and third term in the right-hand side of
Eq.~\eqref{color-decomposition1} vanish. In the last term, the
product of the generators of SU(3) can be simplified using the
following identity:
\begin{align}
t^A_{a'a} t^B_{b'b}=\frac{1}{8}\delta^{AB}t^C_{a'a}
t^C_{b'b}+\left[t^A_{a'a}
t^B_{b'b}-\frac{1}{8}\delta^{AB}t^C_{a'a}t^C_{b'b}\right],\label{colorm}
\end{align}
where only the first term in the right-hand side of
Eq.~\eqref{colorm} survives after multiplying it with the epsilon
tensors $\epsilon_{abc}\epsilon_{a'b'c}$. Then
Eq.~\eqref{color-decomposition1} can be simplified as follows
\begin{align}
\epsilon_{abc}&\epsilon_{a'b'c}(\bar{q}_{a'} \Gamma^\alpha
q_a)(\bar{q}_{b'} \Gamma^{\beta}
q_b)\nonumber\\
=&~\epsilon_{abc}\epsilon_{a'b'c}\bigg\{\frac{1}{9}\delta_{a'a}\delta_{b'b}(\bar{q}
\Gamma^\alpha q)(\bar{q} \Gamma^{\beta}
q)\nonumber\\
&+\frac{1}{2}t^A_{aa'} t^A_{bb'}(\bar{q} \Gamma^\alpha t^B
q)(\bar{q} \Gamma^{\beta} t^B q)\bigg\}\nonumber\\
=&~\frac{2}{3}(\bar{q} \Gamma^\alpha q)(\bar{q} \Gamma^{\beta}
q)-2(\bar{q} \Gamma^\alpha t^B q)(\bar{q} \Gamma^{\beta} t^B q).
\label{color-decomposition}
\end{align}

One can take another Fierz rearrangement to $(\bar{u}\Gamma^{\alpha}
t^A u)(\bar{u}\Gamma^{\beta}  t^A u)$ type of operators in
Eq.~\eqref{4qope1}.  Then one can obtain the following relations
when taking the symmetric and traceless parts of the operator
relations,
\begin{align}
(\bar{u}\gamma^{\alpha} t^A u)(\bar{u}\gamma^{\beta}  t^A
u)\vert_{\textrm{s,t}}=&-\frac{5}{12}(\bar{u}\gamma^{\alpha}
u)(\bar{u} \gamma^{\beta}
u)\vert_{\textrm{s,t}}\nonumber\\
&-\frac{1}{4}(\bar{u} \gamma^{\alpha}\gamma_5 u)(\bar{u}
\gamma^{\beta} \gamma_5
u)\vert_{\textrm{s,t}}\nonumber\\
&+\frac{1}{4}(\bar{u} \sigma_{o}^{~\alpha} u )(\bar{u}
\sigma^{o\beta} u)\vert_{\textrm{s,t}}, \label{cond1}
\end{align}

\begin{align}
(\bar{u}\gamma^{\alpha}\gamma_5 t^A u)(\bar{u}\gamma^{\beta}
\gamma_5 t^A u)\vert_{\textrm{s,t}}=&-\frac{5}{12}(\bar{u}
\gamma^{\alpha}\gamma_5 u)(\bar{u} \gamma^{\beta} \gamma_5
u)\vert_{\textrm{s,t}}\nonumber\\
&-\frac{1}{4}(\bar{u}\gamma^{\alpha} u)(\bar{u} \gamma^{\beta}
u)\vert_{\textrm{s,t}}\nonumber\\
&-\frac{1}{4}(\bar{u} \sigma_{o}^{~\alpha} u )(\bar{u}
\sigma^{o\beta} u)\vert_{\textrm{s,t}}, \label{cond2}
\end{align}
where $\vert_{\textrm{s,t}}$ means symmetric and traceless.
Therefore, only three independent twist-4 (dimension-six spin-2)
matrices remain. Using the twist-4 effects in the deep inelastic
scattering data on the proton and neutron target, one can, in
principle, extract two independent constraints to the three
independent matrix elements. To determine all the matrix elements,
we will additionally use one constraint adopted by
Jaffe~\cite{Jaffe:1981sz}: $(\bar{u} \sigma_{o}^{~\alpha} u
)(\bar{u} \sigma^{o\beta} u)\vert_{\textrm{s,t}}=0$.

\subsubsection{Twist-4 operators with mixed quark flavor}

The second type of four-quark operators appearing in the nucleon sum
rule are of the following-mixed-quark flavor operator form:
\begin{align}
\epsilon_{abc}&\epsilon_{a'bc'}~\gamma^5\gamma^\mu
d_c\bar{d}_{c'}^T\gamma^\nu \gamma^5 ~(u^T_a C\gamma_\mu  \slash
\hspace{-0.2cm}  q \gamma_\nu C \bar{u}^T_{a'})
\nonumber\\
=&~\epsilon_{abc}\epsilon_{a'bc'}~\frac{1}{16}(\gamma^5\gamma^\mu\Gamma_k\gamma^\nu
\gamma^5)(\bar{u}_{a'} \Gamma^o u_{a})(\bar{d}_{c'} \Gamma^k
d_{c})\nonumber\\
&\times\textrm{Tr}\left[\gamma_\mu \slash \hspace{-0.2cm}
q\gamma_\nu C \Gamma_o C\right]\nonumber\\
\Rightarrow&
~\epsilon_{abc}\epsilon_{a'bc'}~\frac{1}{16}\bigg\{-8q_\alpha(\bar{u}_{a'}
\gamma^\alpha u_{a})(\bar{d}_{c'}
d_{c})\nonumber\\
&-8(q_{\beta}\gamma_\alpha+ g_{\alpha\beta}\slash\hspace{-0.2cm} q)(\bar{u}_{a'} \gamma^\alpha u_{a})(\bar{d}_{c'} \gamma^{\beta}  d_{c})\nonumber\\
&+8(q_{\beta}\gamma_\alpha- g_{\alpha\beta}\slash\hspace{-0.2cm}
q)(\bar{u}_{a'} \gamma^\alpha\gamma_5 u_{a})(\bar{d}_{c'}
\gamma^{\beta}\gamma_5 d_{c})\bigg\}, \label{4qope2}
\end{align}
where we have again used Fierz rearrangement to express the
operators in terms of the quark-anti quark type, and have neglected
operators that are odd in parity and time-reversal symmetry.

As in the case with a single quark flavor, the four-quark
condensates in Eq.~\eqref{4qope2} can be decomposed into two
different color structures according to
Eq.~\eqref{color-decomposition} and Eq.~\eqref{colorm}. We cannot
reduce the number of independent operators as in the previous
subsubsection because  performing a similar Fierz rearrangement as
in Eq.~\eqref{cond1} and Eq.~\eqref{cond2}, we find new mixed-flavor
operators of $(\bar{u}\Gamma^{\alpha} d)(\bar{d}\Gamma^{\beta} u)$
type.

\subsubsection{Contributions of dimension-six four-quarks to the OPE}

In summary, the independent four-quark condensates appearing in our nucleon sum
rule are given in Table~\ref{table1}. Not all the matrix elements
are known.

As for the dimension-six spin-0 (scalar) operators, we will assume
the factorized form as $\langle\bar{u}u\rangle^2_{\rho,I}$, although
this assumption has not been justified.  Keeping only the linear
density terms, they can be written as
\begin{align}
\langle[\bar{q}q]_{u,d}\rangle^2_{\rho,I}\Rightarrow&~\langle\bar{q}q\rangle^2_{\textrm{vac}}\nonumber\\
& + 2f\left(1\mp\frac{\mathcal{R}_{-}(m_q)}{\mathcal{R}_{+}(m_q)}
I\right)\langle\bar{q}q\rangle_{\textrm{vac}}\langle[\bar{q}q]_0
\rangle_p\rho,\label{4qscalar}
\end{align}
where $f$ is a parameter introduced in Ref. \cite{Jin:1993up}.

Also, dimension-six spin-1 (vector) operator is factorized up to
linear density terms as in Ref.~\cite{Jin:1993up}.

Dimension-six spin-2 are the twist-4 operators. Th twist-4 operators
appearing in the nucleon sum rule have similar structures as those
appearing in the higher twist effects in deep inelastic
scattering~\cite{Jaffe:1981td,Jaffe:1982pm}. If the higher twist
effects are measured with precision in DIS for the proton and
neutron target, the nucleon expectation value of
$(\bar{u}\gamma^{\alpha}\gamma_5 t^A
u)(\bar{d}\gamma^{\beta}\gamma_5  t^A d)\vert_{\textrm{s,t}}$ can be
estimated with the same precision~\cite{Choi:1993cu}. With further
plausible arguments (Appendix C) on the ratio of $u$ quark and $d$
quark content of the proton such as those used in
Eq.~\eqref{asymmcondensate}, one can estimate the proton expectation
value of $(\bar{u}\gamma^{\alpha} t^A u)(\bar{u}\gamma^{\beta} t^A
u)\vert_{\textrm{s,t}}$,
 $(\bar{u}\gamma^{\alpha}\gamma_5 t^A
u)(\bar{u}\gamma^{\beta}\gamma_5  t^A u)\vert_{\textrm{s,t}}$, and
$(\bar{u}\gamma^{\alpha} t^A u)(\bar{d}\gamma^{\beta}  t^A
d)\vert_{\textrm{s,t}}$.

From these condensates, one can estimate the nucleon expectation
value of all the twist-4 operators for the single flavor case given
in the first column in Table~\ref{table1} with the extra constraint
discussed above. For the mixed-flavor condensates given in the
second column, one cannot deduce all the matrix elements
$(\bar{u}\gamma^{\alpha}\gamma_5 u)(\bar{d}\gamma^{\beta}\gamma_5
d)\vert_{\textrm{s,t}}$ and $(\bar{u}\gamma^{\alpha}
u)(\bar{d}\gamma^{\beta} d)\vert_{\textrm{s,t}}$ from
$(\bar{u}\gamma^{\alpha}\gamma_5 t^A
u)(\bar{d}\gamma^{\beta}\gamma_5  t^A d)\vert_{\textrm{s,t}}$ and
$(\bar{u}\gamma^{\alpha} t^A u)(\bar{d}\gamma^{\beta}  t^A
d)\vert_{\textrm{s,t}}$. We will, however, neglect
$(\bar{u}\gamma^{\alpha}\gamma_5 u)(\bar{d}\gamma^{\beta}\gamma_5
d)\vert_{\textrm{s,t}}$ and $(\bar{u}\gamma^{\alpha}
u)(\bar{d}\gamma^{\beta} d)\vert_{\textrm{s,t}}$ in our present
analysis, as these mixed-quark-flavor condensates do not give
important contributions to the nuclear symmetry energy in the linear
density order.

The proton expectation value of the deducible twist-4 operators can
be parameterized into the following forms:
\begin{align}
\langle(\bar{q}_1&\gamma^{\alpha}\gamma_5 t^A
q_1)(\bar{q}_2\gamma^{\beta} \gamma_5 t^A
q_2)\rangle_p\vert_{\textrm{s,t}}\nonumber\\
=&~\frac{1}{4\pi\alpha_s}\frac{M_N}{2}\left(u^{\alpha}u^\beta
-\frac{1}{4}g^{\alpha\beta}\right)T^1_{q_1q_2},\\
\langle(\bar{q}_1&\gamma^{\alpha} t^A
q_1)(\bar{q}_2\gamma^{\beta}t^A
q_2)\rangle_p\vert_{\textrm{s,t}}\nonumber\\
=&~\frac{1}{4\pi\alpha_s}\frac{M_N}{2}\left(u^{\alpha}u^\beta -\frac{1}{4}g^{\alpha\beta}\right)T^2_{q_1q_2},\\
\langle(\bar{q}_1&\gamma^{\alpha}\gamma_5
q_1)(\bar{q}_2\gamma^{\beta} \gamma_5
q_2)\rangle_p\vert_{\textrm{s,t}}\nonumber\\
=&~\frac{1}{4\pi\alpha_s}\frac{M_N}{2}\left(u^{\alpha}u^\beta -\frac{1}{4}g^{\alpha\beta}\right)T^3_{q_1q_2},\\
\langle(\bar{q}_1&\gamma^{\alpha} q_1)(\bar{q}_2\gamma^{\beta}
q_2)\rangle_p\vert_{\textrm{s,t}}\nonumber\\
=&~\frac{1}{4\pi\alpha_s}\frac{M_N}{2}\left(u^{\alpha}u^\beta
-\frac{1}{4}g^{\alpha\beta}\right)T^4_{q_1q_2},
\end{align}
where `$q_1$' and $`q_2$' represent quark flavors. We have extracted
the $T^i$s from the matrix elements estimated in
Ref.~\cite{Choi:1993cu} and listed in Table~\ref{table2}.

Using the parametrization of the nucleon expectation value of the
twist-4 operators together with the linear density approximation
given in Eq.~\eqref{condensatea}, the contributions to the
correlation function from the four-quark operators can be written as

\begin{widetext}\allowdisplaybreaks

\begin{table}
\begin{tabular}{c c c c c}
\hline\hline  Quark flavor &&$q_1=q_2=q$ & & $q_1\neq q_2$
\\
\hline && $(\bar{q}\gamma^{\alpha}\gamma_5 q)(\bar{q}\gamma^{\beta}
\gamma_5 q)\vert_{\textrm{s,t}}$ & &
$(\bar{q}_{1}\gamma^{\alpha}\gamma_5
q_{1})(\bar{q}_{2}\gamma^{\beta} \gamma_5
q_{2})\vert_{\textrm{s,t}}$  \\
$\textrm{Dimension-six spin-2}$ & & $(\bar{q}\gamma^{\alpha}
q)(\bar{q}\gamma^{\beta} q)\vert_{\textrm{s,t}}$ &
&$(\bar{q}_{1}\gamma^{\alpha} q_{1})(\bar{q}_{2}\gamma^{\beta}
q_{2})\vert_{\textrm{s,t}}$  \\
($\textrm{twist-4}$)&&$(\bar{q} \sigma_{o}^{~\alpha}
q)(\bar{q}\sigma^{o\beta} q)\vert_{\textrm{s,t}}$ & &
$(\bar{q}_{1}\gamma^{\alpha}\gamma_5 t^A
q_{1})(\bar{q}_{2}\gamma^{\beta} \gamma_5 t^A
q_{2})\vert_{\textrm{s,t}}$  \\
 & &   &
&$(\bar{q}_{1}\gamma^{\alpha} t^A q_{1})(\bar{q}_{2}\gamma^{\beta}
t^A
q_{2})\vert_{\textrm{s,t}}$  \\
\\ $\textrm{Dimension-six spin-1}$&& &  &$(\bar{q}_{1} \gamma^\alpha
q_{1})(\bar{q}_{2}
q_{2})$\\
($\textrm{vector}$)&& &  &$(\bar{q}_{1} \gamma^\alpha t^A
q_{1})(\bar{q}_{2} t^A
q_{2})$\\
\\ && $(\bar{q}\gamma_{\alpha}\gamma_5 q)(\bar{q}\gamma^{\alpha}
\gamma_5 q)$ & & $(\bar{q}_{1}\gamma_{\alpha}\gamma_5
q_{1})(\bar{q}_{2}\gamma^{\alpha} \gamma_5
q_{2})$  \\
$\textrm{Dimension-six spin-0}$ & & $(\bar{q}\gamma_{\alpha}
q)(\bar{q}\gamma^{\alpha} q)$ & &$(\bar{q}_{1}\gamma_{\alpha}
q_{1})(\bar{q}_{2}\gamma^{\alpha}
q_{2})$  \\
($\textrm{scalar}$)&&$(\bar{q}\sigma_{o\alpha} q)(\bar{q}
\sigma^{o\alpha} q)$ & & $(\bar{q}_{1}\gamma_{\alpha}\gamma_5 t^A
q_{1})(\bar{q}_{2}\gamma^{\alpha} \gamma_5 t^A
q_{2})$  \\
 & &  & &$(\bar{q}_{1}\gamma_{\alpha}
t^A q_{1})(\bar{q}_{2}\gamma^{\alpha} t^A
q_{2})$ \\
\hline\hline
\end{tabular}
\caption{Independent four-quark operators appearing in the nucleon
OPE with Ioffe's interpolating current. `$q_1$' and `$q_2$'
represent light quark flavors.}\label{table1}
\end{table}

\begin{table}
\centering
\begin{tabular}{l c c c c c c c c c c}
\hline\hline   &$T^1_{uu}$ & $T^1_{dd}$ & $T^2_{uu}$ & $T^2_{dd}$ &
$T^3_{uu}$ & $T^3_{dd}$ & $T^4_{uu}$ & $T^4_{dd}$ & $T^1_{ud}$
&$T^2_{ud}$
\\ \hline &&&&First set&&&&\\

$K^1_u=K^1_{ud}/\beta$&-0.132 & -0.041 & 0.154 & 0.048 & 0.842 & 0.262 & -0.875 & -0.272 & -0.042 & 0.049 \\
$K^1_u=K^1_{ud}(\beta+1)/\beta$&-0.071 & -0.012 & 0.070 & 0.012 & 0.424 & 0.072 & -0.422 & -0.072 & -0.042 & 0.041 \\
$K^1_u=K^1_{ud}$&-0.042 & 0.002 & 0.033 & -0.002 & 0.240 & -0.012 & -0.233 & 0.012 & -0.042 & 0.031 \\
  &&&&Second set&&& \\
$K^1_u=-K^1_{ud}$& 0.215 & 0.124& -0.432 & -0.265 & -1.778 & -1.091 & 2.104 & 1.290 & -0.042 & 0.057 \\
$K^1_u=-K^1_{ud}(\beta+1)/\beta$& 0.154 & 0.100 & -0.337 & -0.219 & -1.336 & 0.868 & 1.610 & 1.046 & -0.042 & 0.056 \\
$K^1_u=-K^1_{ud}/\beta$& 0.125 & 0.085 & -0.297 & -0.202 & -1.137 & -0.773 & 1.395 & 0.949 & -0.042 &  0.058 \\
\hline \hline
\end{tabular}\caption{Two sets for $T^i$s. The three different classifications of
$T^i$ follow that given in Ref.~\cite{Choi:1993cu}. Detailed
treatment is given in Appendix C. Units are in
$\textrm{GeV}^2$.}\label{table2}
\end{table}

\begin{align}
\Pi_{D=6,s}^O(q_0^2,\vert\vec{q}\vert)
=&-\frac{4}{3}\frac{1}{q^2}\langle\bar{q}q\rangle_{\textrm{vac}}\left(\frac{3}{2}-\frac{1}{2}I\right)\rho,\\
\Pi_{D=6,q}^E(q_0^2,\vert\vec{q}\vert)
=&-\frac{2}{3q^2}\langle\bar{u}u\rangle^2_{\rho,I}+\frac{1}{q^2}\frac{1}{4\pi\alpha_s}\frac{M_N}{2}
[T^1_{ud}-T^2_{ud}]
\rho+\frac{1}{q^2}\frac{1}{4\pi\alpha_s}\frac{M_N}{2}
\left([T^1_{~0}-T^2_{~0}]-[T^1_{~1}-T^2_{~1}]I\right)\rho\nonumber\\
&-\frac{1}{3q^2}\frac{1}{4\pi\alpha_s}\frac{M_N}{2}
\left([T^3_{~0}-T^4_{~0}]-[T^3_{~1}-T^4_{~1}]I\right)\rho,
\\
\Pi_{D=6,u}^O(q_0^2,\vert\vec{q}\vert)=&-\frac{4}{q^2}\frac{1}{4\pi\alpha_s}\frac{M_N}{2}
 [T^1_{ud}-T^2_{ud}] \rho-\frac{4}{q^2}\frac{1}{4\pi\alpha_s}\frac{M_N}{2}
\left([T^1_{~0}-T^2_{~0}]-[T^1_{~1}-T^2_{~1}]I\right)\rho\nonumber\\
 &+\frac{4}{3q^2}\frac{1}{4\pi\alpha_s}\frac{M_N}{2}
\left([T^3_{~0}-T^4_{~0}]-[T^3_{~1}-T^4_{~1}]I\right)\rho,
\end{align}
where $T^i_{~0}=\frac{1}{2}(T^i_{uu}+T^i_{dd})$ and
$T^i_{~1}=\frac{1}{2}(T^i_{uu}-T^i_{dd})$.
All the vacuum and
scalar condensates are factorized as Eq.~\eqref{factorization} and
Eq.~\eqref{4qscalar}. The corresponding Borel transformations are
given as follows:
\begin{align}
\bar{\mathcal{B}}[\Pi_{D=6,s}(q_0^2,\vert\vec{q}\vert)]=
&~(-\bar{E}_q) \frac{4}{3}\langle\bar{q}q\rangle_{\textrm{vac}}\left(\frac{3}{2}-\frac{1}{2}I\right)\rho,\\
\bar{\mathcal{B}}[\Pi_{D=6,q}(q_0^2,\vert\vec{q}\vert)]=
&~\frac{2}{3}\langle\bar{u}u\rangle^2_{\rho,I}
L^{\frac{4}{9}}-\frac{1}{4\pi\alpha_s}\frac{M_N}{2}\bigg\{
 [T^1_{ud}-T^2_{ud}]
+ \left([T^1_{~0}-T^2_{~0}]-[T^1_{~1}-T^2_{~1}]I\right)\nonumber\\
&-\frac{1}{3}
\left([T^3_{~0}-T^4_{~0}]-[T^3_{~1}-T^4_{~1}]I\right)\bigg\}\rho
L^{-\frac{4}{9}},\\
\bar{\mathcal{B}}[\Pi_{D=6,u}(q_0^2,\vert\vec{q}\vert)] =&~
\frac{(4\bar{E}_q)}{4\pi\alpha_s}\frac{M_N}{2}\left\{
 [T^1_{ud}-T^2_{ud}] +
\left([T^1_{~0}-T^2_{~0}]-[T^1_{~1}-T^2_{~1}]I\right)-\frac{1}{3}
\left([T^3_{~0}-T^4_{~0}]-[T^3_{~1}-T^4_{~1}]I\right)\right\}\rho
L^{-\frac{4}{9}}.
\end{align}
\end{widetext}

Here, we have neglected the scaling of the matrix elements coming
from  the anomalous dimension of the dimension-six  operators
$\Gamma_{O_n}$. Although the twist-4 matrix elements are estimated
at the separation scale of 5~GeV, and the matrix element we need is
at lower energy scale close to the Borel mass, we will neglect the
running of the matrix elements through the anomalous dimension for
operator $\Gamma_{O_n}$, because the present estimate of the matrix
elements already contains $\pm50\%$ uncertainty. Throughout this
paper, we used $\alpha_s\simeq0.5$ for these twist-4 matrix elements
as in Ref.~\cite{Choi:1993cu,Jaffe:1981td}. In principle, the
coupling appearing in the twist-4 matrix element should run with the
Borel mass.  However we neglect such running because within the
region of Borel mass $1.0~\textrm{GeV}^2\leq M^2
\leq1.2~\textrm{GeV}^2$, $\alpha_s(M^2) \sim 0.4$ and, hence, the
difference with what was used is within the uncertainty of the
twist-4 matrix element.

\section{Results for the nucleon sum rule and the nuclear symmetry energy}
We have expressed the self-energy contributions of the nucleons that
contribute to the nucleon energy as Eq.~\eqref{polep} in terms of
the Borel transformed OPE as given in Eqs.~\eqref{borel1},
\eqref{borel2}, and \eqref{borel3}. The next step is to substitute
Eq.~\eqref{polep} to Eq.~\eqref{define2} to extract the symmetry
energy as defined in Eq.~\eqref{def-sym}. There then will be the
trivial kinematic correction coming from the three-momentum
dependence in the kinetic energy part of Eq.~\eqref{polep}. This
term is universal and corresponds to the term in
Eq.~\eqref{kinetic}. Instead of following the full procedure, in
this work, we will just concentrate on the contribution coming from
the scalar and vector self-energy. This corresponds to calculating
the contribution to the nuclear symmetry energy from potentials in
effective models.

\subsection{QCD sum rule formula}

The quasi nucleon self-energies in the rest frame can be obtained in
QCD sum rules by taking the ratios
Eq.~\eqref{borel1}/Eq.~\eqref{borel2} and
Eq.~\eqref{borel3}/Eq.~\eqref{borel2} for both the proton and
neutron as follows:
\begin{align}
E_{q,V(I)} & \equiv \Sigma_v+M_N^{*} = \frac{\mathcal{N}^{n,p}(\rho)}{\mathcal{D}^{n,p}(\rho)} \nonumber \\
& =\frac{\bar{\mathcal{B}}[\Pi^{n,p}_s(q_0^2,\vert\vec{q}\vert)]
+
\bar{\mathcal{B}}[\Pi^{n,p}_u(q_0^2,\vert\vec{q}\vert)]}{\bar{\mathcal{B}}[\Pi^{n,p}_q(q_0^2,\vert\vec{q}\vert)]},
\end{align}
where subscripts  $q,V(I)$ are meant to represent the potential part
of Eq.~(\ref{polep}) in the asymmetric nuclear matter. To discuss
different approximations of self-energies in terms of the density
and the asymmetric factor, we introduce the following symbols
$\mathcal{N}^{n,p}_{(\rho^m,I^l)}(\rho)$ and
$\mathcal{D}^{n,p}_{(\rho^m,I^l)}(\rho)$:
\begin{align}
\mathcal{N}^{n,p}(\rho) = &~ \mathcal{N}^{n,p}_{(\rho^0,I^0)} +
\mathcal{N}^{n,p}_{(\rho,I^0)}\rho+ \left[\mathcal{N}^{n,p}_{(\rho,I)}\rho\right]I\nonumber\\
&+\sum_2^m\sum_2^l \left[\mathcal{N}^{n,p}_{(\rho^m,I^l)}\rho^m \right]I^l,\\
\mathcal{D}^{n,p}(\rho)= &~ \mathcal{D}^{n,p}_{(\rho^0,I^0)} +
\mathcal{D}^{n,p}_{(\rho,I^0)} \rho  +
\left[\mathcal{D}^{n,p}_{(\rho,I)}\rho\right]I\nonumber\\
&+\sum_2^m\sum_2^l \left[\mathcal{D}^{n,p}_{(\rho^m,I^l)}\rho^m
\right]I^l,
\end{align}
where the superscripts $n$ and $p$ represent either the neutron or
the proton, respectively. For the pair of subscripts $(\rho^m,I^l)$,
the first index represents the order of the density, while the
second index represents the isospin. Due to isospin symmetry, the
isoscalar terms have the following relations:
\begin{align}
\mathcal{N}^{n}_{(\rho^m,I^l)}&=(-1)^l\mathcal{N}^{p}_{(\rho^m,I^l)},\\
\mathcal{D}^{n}_{(\rho^m,I^l)}&=(-1)^l\mathcal{D}^{p}_{(\rho^m,I^l)},
\end{align}
where $l$ is the integer for the order of the isospin.  All these
terms are summarized in Appendix D.

Because the dominant term of $\mathcal{D}^{n,p}(\rho)$ is
$\mathcal{D}^{n,p}_{(\rho^0,I^0)}$, one can expand the denominator
in terms of $(1/\mathcal{D}^{n,p}_{(\rho^0,I^0)})$ times condensate.
After rewriting this with powers of $\rho$ and $I$, one can express
the potential part of a single nucleon energy as
\begin{align}
E^{n,p}_V(\rho,I) &=
E^{n,p}_{V,(\rho^0,I^0)}+\sum_{k=1}^{\infty}\sum_{i=0}^{\infty}\left(
\left[E^{n,p}_{V,(\rho^k,I^i)}\rho^k\right]I^i\right),\label{self-ope}
\end{align}
where $E^{n,p}_{V,(\rho^k,I^i)}$ are written in terms of
$\mathcal{N}^{n,p}(\rho)$ and $\mathcal{D}^{n,p}(\rho)$. Averaging
Eq.~\eqref{self-ope} as Eq.~\eqref{define2} and collecting terms of
$I^2$ from Eq.~\eqref{define}, one can extract
$E^{\textrm{sym}}_V(\rho)$ as follows:

\begin{align}
E^{\textrm{sym}}_V(\rho)=&~\frac{1}{2}\bigg[
\frac{1}{2}\rho\left(E^{n}_{V,(\rho,I)}-E^{p}_{V,(\rho,I)}\right)
\nonumber\\
&+\frac{1}{3}\rho^2\left(E^{n}_{V,(\rho^2,I)}-E^{p}_{V,(\rho^2,I)}\right)
\nonumber\\
&+\frac{1}{4}\rho^3\left(E^{n}_{V,(\rho^3,I)}-E^{p}_{V,(\rho^3,I)}\right)+ \cdots \bigg] \nonumber\\
&+ \frac{1}{2}\bigg[
\frac{1}{3}\rho^2\left(E^{n}_{V,(\rho^2,I^2)}+E^{p}_{V,(\rho^2,I^2)}\right)
\nonumber\\&+\frac{1}{4}\rho^3\left(E^{n}_{V,(\rho^3,I^2)}+E^{p}_{V,(\rho^3,I^2)}\right)+\cdots\bigg].\label{densityexpansion}
\end{align}

For terms linear in density, one can see that the first term in the
upper bracket of Eq.~\eqref{densityexpansion} corresponds to the
form given in Eq.~\eqref{sym2}. The explicit expression in terms of
$\mathcal{N}^{n}_{(\rho^m,I^l)}$ and
$\mathcal{D}^{n}_{(\rho^m,I^l)}$ is
\begin{align}
E^{\textrm{sym}}_{V,\rho}=\frac{1}{4}\rho
\bigg[\frac{1}{\mathcal{D}^{p}_{(\rho^0,I^0)}}(-2\mathcal{N}^{p}_{(\rho,I)})
-\frac{\mathcal{N}^{p}_{(\rho^0,I^0)}}{(\mathcal{D}^{p}_{(\rho^0,I^0)})^2}
(-2\mathcal{D}^{p}_{(\rho,I)})\bigg],\label{linearexpansion}
\end{align}
valid to leading order in density.

When higher density dependence of the condensates is calculated,
Eq.~\eqref{densityexpansion} provides a systematic expression of
$E^{\textrm{sym}}_V(\rho)$ that includes higher $\rho^{n\geq2}$
terms.

\subsection{Sum rule analysis}

In principle, a physical quantity extracted from the QCD sum rule
should not depend on the Borel parameter $M^2$. However, since we
truncate the OPE at finite mass dimension, such a physical quantity
should be obtained within a reliable range of $M^2$ (Borel window)
with ``plateau''. While we do not find the most stable a ``plateau''
with an extremum in the appropriate Borel window, one finds that the
results have only a weak dependence on $M^2$.

The well accepted Borel window for the nucleon sum rule is $0.8~
\textrm{GeV}^2\leq M^2 \leq 1.4~\textrm{GeV}^2$~\cite{Ioffe:1983ju}.
But as our sum rule contains the newly added twist-4 four-quark
operators, the Borel window needs to be re-examined. We determine
the upper Borel window by requiring that the quasi nucleon
contribution is more than 50$\%$ of the total sum rule so the
continuum contribution is less than 50$\%$. As for the lower limit,
for the same OPE, we restrict the contribution from the highest mass
dimension operator to be less than 50$\%$ of the total contribution.
For the quasi nucleon energy in medium rest frame, we applied this
prescription to the right-hand side of Eqs.~\eqref{borel1},
\eqref{borel2}, and \eqref{borel3}.

\begin{figure}
\includegraphics[width=4.22cm,height=5.125cm]{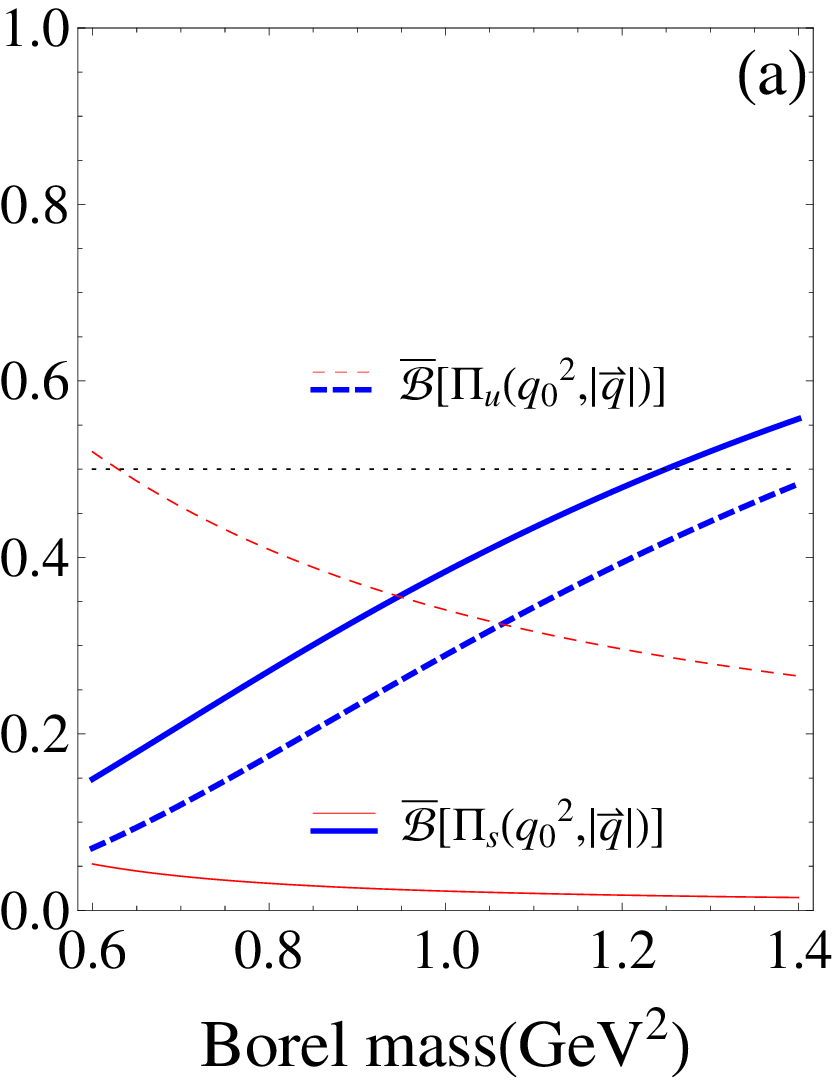}
\includegraphics[width=4.22cm,height=5.125cm]{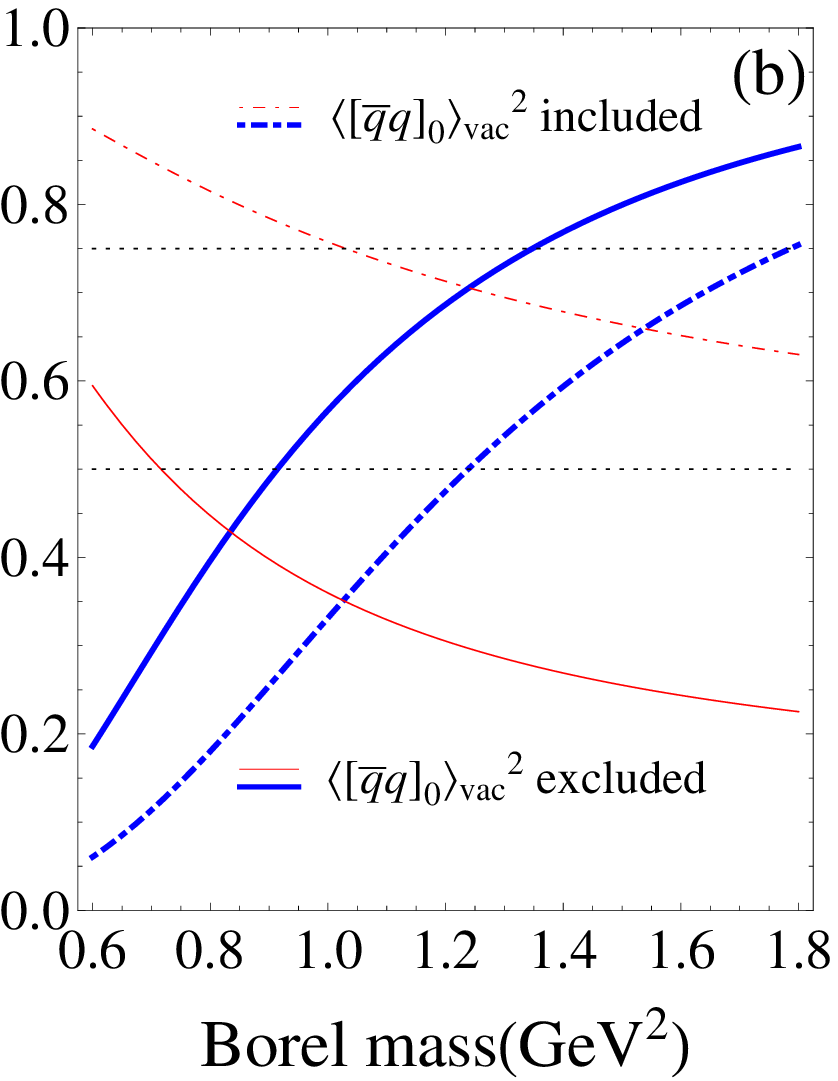}
\caption{(Color online) Borel window for (a)
$\bar{\mathcal{B}}[\Pi_s(q_0^2,\vert\vec{q}\vert)]$ and
$\bar{\mathcal{B}}[\Pi_u(q_0^2,\vert\vec{q}\vert)]$, and (b)
$\bar{\mathcal{B}}[\Pi_q(q_0^2,\vert\vec{q}\vert)]$. In both
figures, the thick lines increasing with the Borel mass represent
the ratio (the contribution of highest dimensional operators)/(the
total OPE), and the thin lines decreasing with the Borel mass
represent the ratio (the continuum contribution)/(total
contribution). These graphs are obtained with $T^i$s in the
$K^1_u=K^1_{ud}(\beta+1)/\beta$ estimation from the first set of
Table~\ref{table2}.}\label{borelwindow}
\end{figure}

The Borel curves for the three invariants [Eqs.~\eqref{borel1},
\eqref{borel2}, and \eqref{borel3}] are plotted in
Fig.~\ref{borelwindow}. Here, all the graphs are obtained with the
$T^i$s using the $K^1_u=K^1_{ud}(\beta+1)/\beta$ estimates from the
first set of Table~\ref{table2}. From Fig.~\ref{borelwindow}(a), one
can get acceptable Borel windows for
$\bar{\mathcal{B}}[\Pi_s(q_0^2,\vert\vec{q}\vert)]$
[Eq.~\eqref{borel1}] and
$\bar{\mathcal{B}}[\Pi_u(q_0^2,\vert\vec{q}\vert)]$
[Eq.~\eqref{borel3}].  However, in Fig.~\ref{borelwindow}(b),
$\bar{\mathcal{B}}[\Pi_q(q_0^2,\vert\vec{q}\vert)]$ do not provide
an acceptable Borel window. While the usual Borel window is obtained
by requiring that the power and continuum corrections are both less
than 50$\%$ of the total OPE, we will loosen the condition to be
less than 75$\%$ in this case.

This large power correction may be caused by an overestimated
$\langle [\bar{q}q]_0\rangle^2_{\textrm{vac}}$. As mentioned in the
previous section, all the vacuum expectation values of four-quark
operators are factorized as in Eq.~(\ref{4qscalar}). Only large
$N_c$ supports factorization in the vacuum.  Hence, the
generalization to the nuclear medium can be only an order of
magnitude estimate with large uncertainty.   For example, as one can
see in Fig.~\ref{borelwindow}(b), the lower and upper boundaries
from $\bar{\mathcal{B}}[\Pi_q(q_0^2,\vert\vec{q}\vert)]$ are already
largely affected by whether the vacuum value  $\langle
[\bar{q}q]_0\rangle^2_{\textrm{vac}}$ is included. Another reason
for the larger uncertainty could be the neglected twist-4 matrix
elements $T^3_{ud}$ and $T^4_{ud}$ for
$(\bar{u}\gamma^{\alpha}\gamma_5 u)(\bar{d}\gamma^{\beta}\gamma_5
d)\vert_{\textrm{s,t}}$ and $(\bar{u}\gamma^{\alpha}
u)(\bar{d}\gamma^{\beta} d)\vert_{\textrm{s,t}}$. If the vacuum
expectation value of four-quark operators as well as the $T^3_{ud}$
and $T^4_{ud}$ can be determined well, we can discuss about the
stability of our sum rule in a more reliable way. The second set of
$T^i$s from Table~\ref{table2} do not produce any acceptable Borel
window. In conclusion, we will use the results from the following
Borel window: $1.0~\textrm{GeV}^2\leq M^2 \leq 1.2 ~\textrm{GeV}^2$.

In the analysis to follow, for the symmetric nuclear matter case, we
will denote the twist-4 condensates contribution to the quasi
nucleon self-energy as $\Sigma_T$ and the total quasi nucleon
self-energy in the rest frame as $E_{q,V(I=0)}$. For the asymmetric
nuclear matter case, we will use two sum rules for
$E^{\textrm{sym}}_V$: one that includes contributions up to order
$\rho$ terms
 and another one up to $\rho^2$. The former sum rule will be called the
linear $\rho$ sum rule ($E^{\textrm{sym}}_{V,\rho}$) and the latter
the $\rho^{2}$ sum rule ($E^{\textrm{sym}}_{V,\rho^2}$). As for the
value for the anti nucleon pole,  an optimal ``in-medium'' value
ranged $-0.2~\textrm{GeV}\leq\bar{E}_q\leq-0.4~\textrm{GeV}$ will be
used for each different estimation of twist-4 matrix elements in the
sum rule for the quasi nucleon self-energy, while the ``bare'' value
$\bar{E}_q=-M_N$ will be used in the sum rule for the nuclear
symmetry energy. This is so because the quasi hole contribution in
the nuclear symmetry energy comes with a term proportional to the
density [Eq.~\eqref{define2}]. Nuclear matter density $\rho$ is set
at the saturation density $\rho_0=0.16~\textrm{fm}^{-3}$ and the
corresponding quasi nucleon three-momentum $\vert\vec{q}\vert$ is
taken to be $270~\textrm{MeV}$, the Fermi momentum of a normal
nucleus ($\rho_0=0.16~\textrm{fm}^{-3}$). The light quark ($u$, $d$
quark) mass $m_q$ is taken to be $5~\textrm{MeV}$.

As for the density dependence of dimension-six spin-0 condensates,
different $f$ values are used for every estimation of twist-4 matrix
elements.  For the first set of Table~\ref{table2}; $f=-0.2$ for
$K^1_u=K^1_{ud}/\beta$ (corresponding
$\bar{E}_q=-0.26~\textrm{GeV}$), $f=-0.12$ (corresponding
$\bar{E}_q=-0.30~\textrm{GeV}$)for $K^1_u=K^1_{ud}(\beta+1)/\beta$
and, $f=-0.08$ for $K^1_u=K^1_{ud}$ (corresponding
$\bar{E}_q=-0.34~\textrm{GeV}$). This parameter set of $f$'s are
chosen to satisfy the self consistency constraint as given in
Eq.~\eqref{pole} for the quasi hole value.  Again, the second set in
Table~\ref{table2} does not provide a set of $f$'s which satisfies
the constraint of Eq.~\eqref{pole}. A detailed discussion for
related parameters ($f$ and $\bar{E}_q$) will be given in a later
section.

\subsubsection{Symmetric nuclear matter}

\begin{figure}
\includegraphics[width=4.27cm,height=5.125cm]{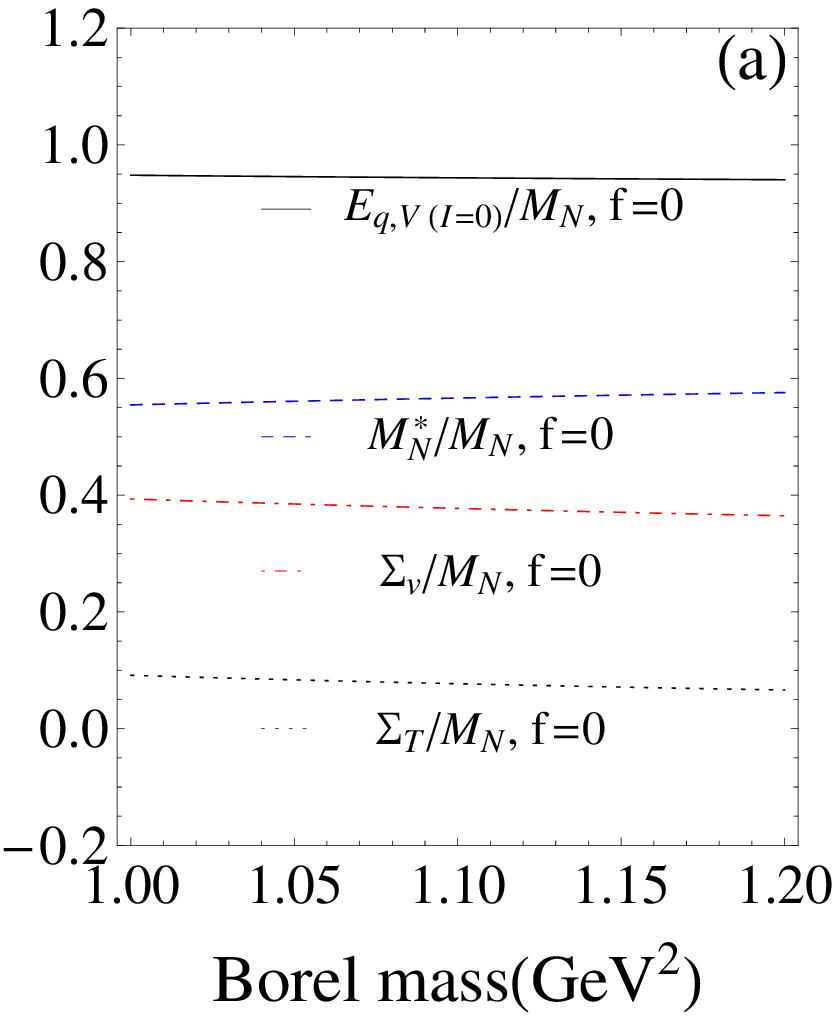}
\includegraphics[width=4.27cm,height=5.125cm]{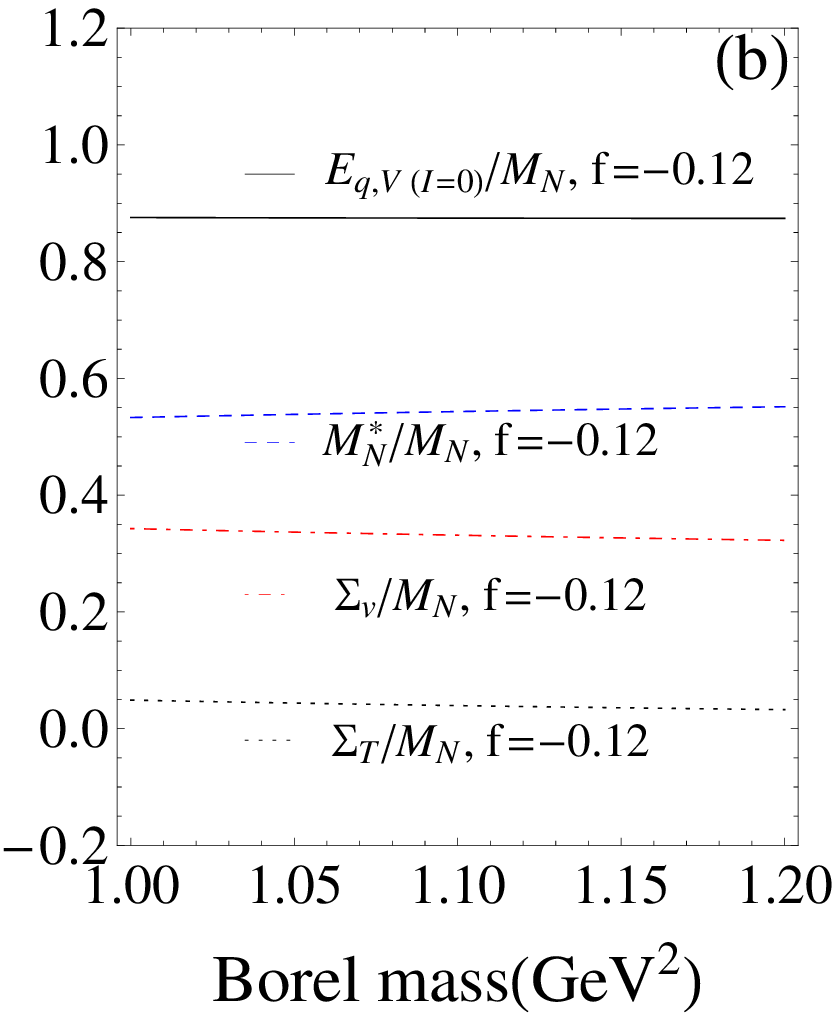}
\caption{(Color online) The ratios between quasi nucleon self
energies and the vacuum mass.  The different lines represent
[$M^{*}_N/M_N$ (dashed blue), $\Sigma_v/M_N$ (dot-dashed red),
$\Sigma_T/M_N$ (dotted black) and $E_{q,V(I=0)}/M_N$ (solid black)],
respectively.}\label{nucla}
\end{figure}

First, we investigate the quasi nucleon self-energies in the
symmetric nuclear matter with twist-4 condensates. Throughout the
analysis, we check the result against the $I=0$ case. In
Fig.~\ref{nucla}, we plot the ratio to the nucleon mass in a vacuum
of the in-medium scalar self-energy ($M^{*}_N/M_N$), the vector
self-energy ($\Sigma_v/M_N$), the twist-4 condensate contribution
($\Sigma_T/M_N$), and the potential part of the total quasi nucleon
self-energy in the rest frame [$E_{q,V(I=0)}/M_N$]. For the  twist-4
matrix elements, we take $K^1_u=K^1_{ud}(\beta+1)/\beta$ from the
first set in Table~\ref{table2} and the corresponding
$\bar{E}_q=-0.30$ $\textrm{GeV}$, which gives the average result.
From our analysis, we find that the contribution of the the twist-4
condensates give enhancement of the quasi nucleon self-energy by
$\sim50~\textrm{MeV}$. When $f=0$, we find the ratio
$E_{q,V(I=0)}/M_N\simeq0.96$, $M^{*}_N/M_N\simeq0.58$ and
$\Sigma_v/M_N\simeq0.37$. By using the aforementioned parameter set
for $f<0$ and $\bar{E}_q$, the ratios become
$E_{q,V(I=0)}/M_N\simeq0.87$, $M^{*}_N/M_N\simeq0.56$ and
$\Sigma_v/M_N\simeq0.30$, which are comparable with previous
studies~\cite{Cohen:1991js,Furnstahl:1992pi,Jin:1992id}. When the
second set of Table~\ref{table2} is used for the twist-4 matrix
elements, we do not find a stable behavior in the same Borel window
in contrast to the case with the first set as shown previously. By
setting $f>0$ for the second set, $E_{q,V(I=0)}/M_N$ can be adjusted
to $\sim0.9$ which is a typically acceptable value. But even so,
there is no reasonable $f$ and $\bar{E}_q$ for $M^{*}_N/M_N$ and
$\Sigma_v/M_N$ which satisfies Eq.~\eqref{pole}. The estimates for
$T^i$s given in the second set of Table~\ref{table2} do not
reproduce the aspect of the nucleon sum rule that is consistent with
the Dirac Phenomenology \cite{Cohen:1991js}. Hence, we will continue
the present analysis with estimates for $T^i$s given by the the
first set in Table~\ref{table2}.

\begin{figure}[b]
\includegraphics[width=4.27cm,height=5.125cm]{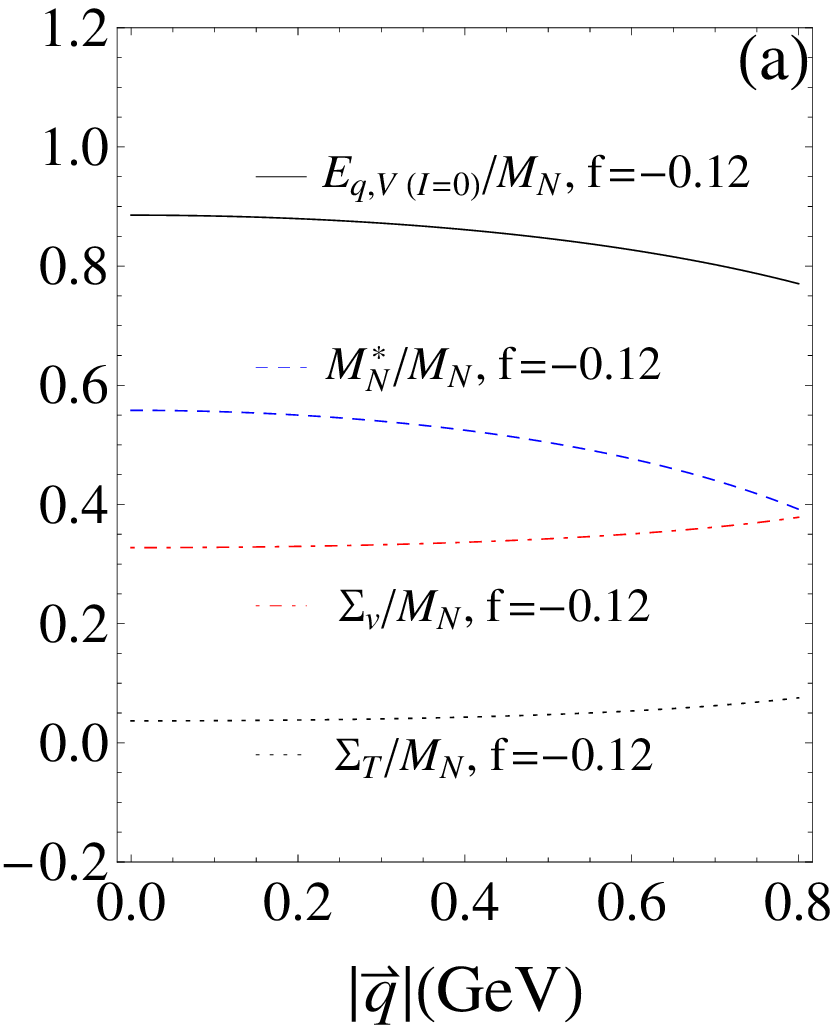}
\includegraphics[width=4.27cm,height=5.125cm]{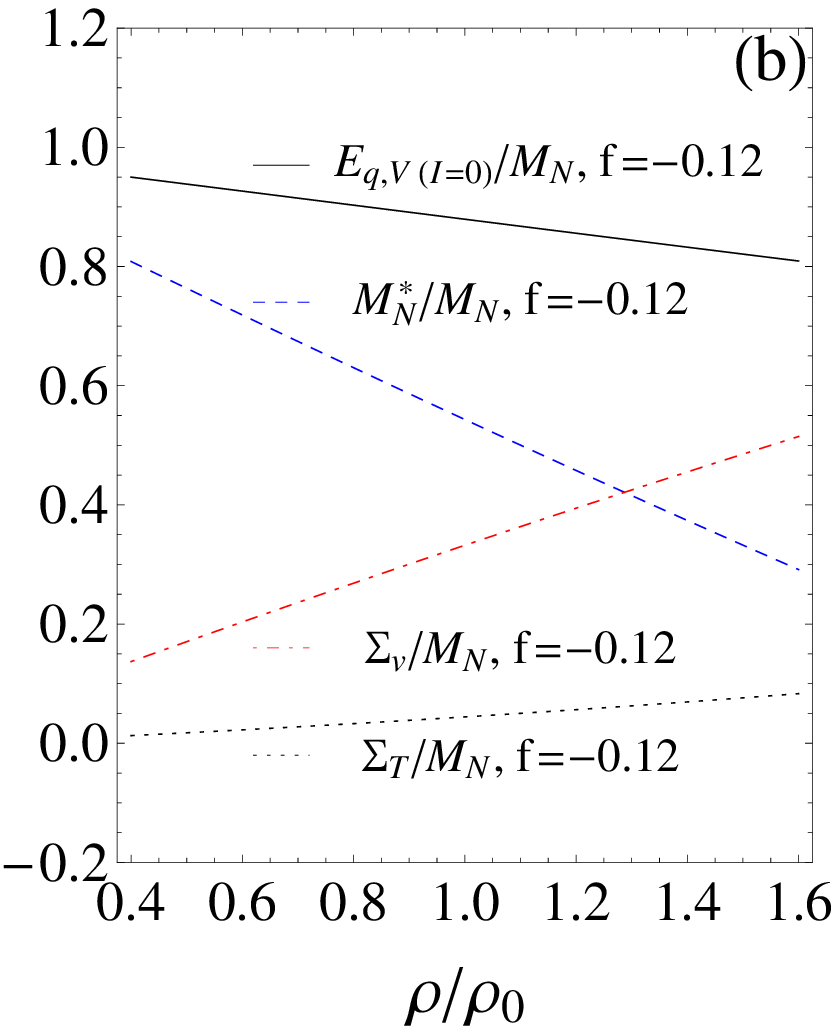}
\caption{(Color online) (a) $\vert \vec{q} \vert$ and (b) density
dependence of the ratios between quasi nucleon self-energies and the
vacuum mass.} \label{nuclb}
\end{figure}

The quasi nucleon three-momentum dependence is plotted in
Fig.~\ref{nuclb}(a)  for the $f<0$ case; one finds that the ratios
$\Sigma_v/M_N$ and $\Sigma_T/M_N$ do not depend strongly on the
quasi nucleon three-momentum.  On the other hand, $M^{*}_N/M_N$
shows a significant change when $\vert\vec{q}\vert\geq0.5$
$\textrm{GeV}$ as in Ref.~\cite{Jin:1993up}. So this sum rule
analysis works in the $0\leq\vert\vec{q}\vert\leq0.5$ $\textrm{GeV}$
region, which is consistent with our phenomenological ansatz that
assumes a momentum-independent self-energy.

As all the condensates in our nucleon sum rule are estimated to
linear order in density, the results may be valid at least near the
nuclear saturation density region. In Fig.~\ref{nuclb}(b), the
density dependence of the quasi nucleon self-energies is plotted for
$0.4\leq\rho/\rho_0\leq 1.6$. Here we used the parameter set
$f=-0.12$ and $\bar{E}_q=-0.30$ $\textrm{GeV}$ determined at the
saturation density, as our nucleon sum rule do not depend strongly
on $\bar{E}_q$ as long as it is varied within
$-0.6~\textrm{GeV}\leq\bar{E}_q\leq-0.3~\textrm{GeV}$ which covers
the naive estimates for  $\bar{E}_q$ when
$0.4\leq\rho/\rho_0\leq1.6$.  One also notes that the magnitude of
both $\Sigma_v/M_N$ and $\Sigma_T/M_N$ increases with density while
$M^{*}_N/M_N$ reduces.

\subsubsection{Asymmetric nuclear matter}

In our sum rule, the nuclear bulk properties in the asymmetric
nuclear matter are parameterized by the asymmetry factor $I$. If one
plots the quasi nucleon self-energy as a function of $I$ to leading
order in density, $E^{\textrm{sym}}_{V,\rho}$ can be obtained from
the difference between the slopes of the quasi neutron and the quasi
proton [Eq.~\eqref{linearexpansion}].

\begin{figure}
\includegraphics[width=4.27cm,height=5.125cm]{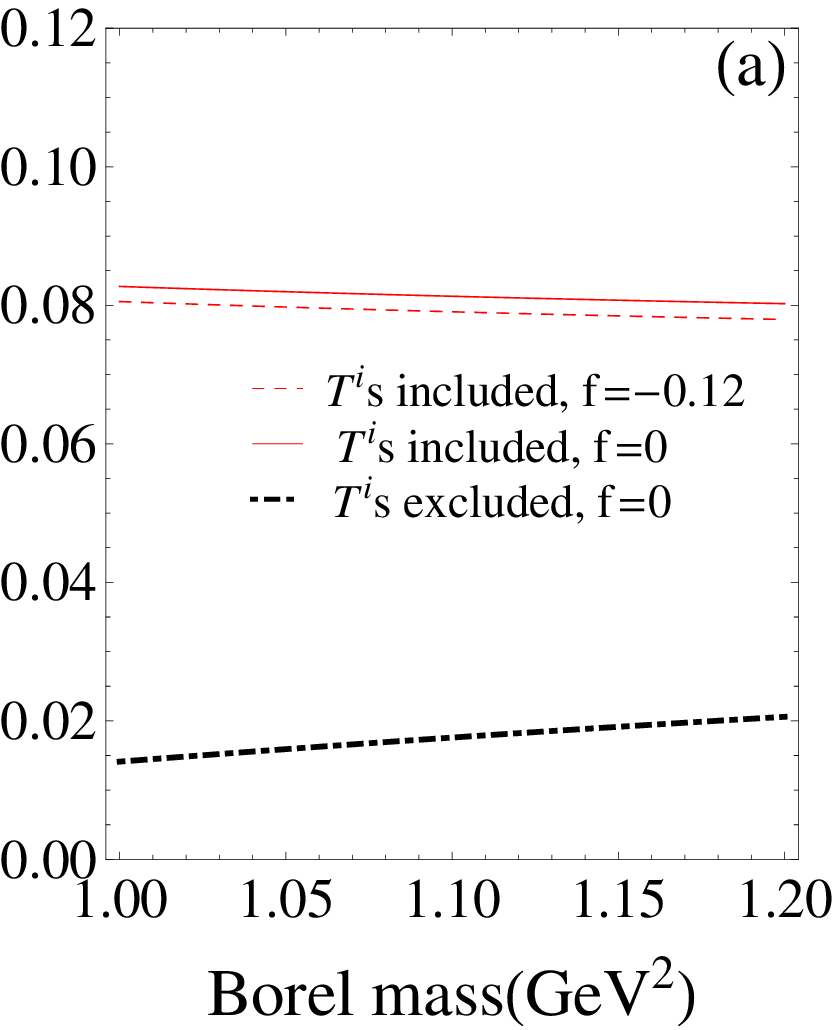}
\includegraphics[width=4.27cm,height=5.125cm]{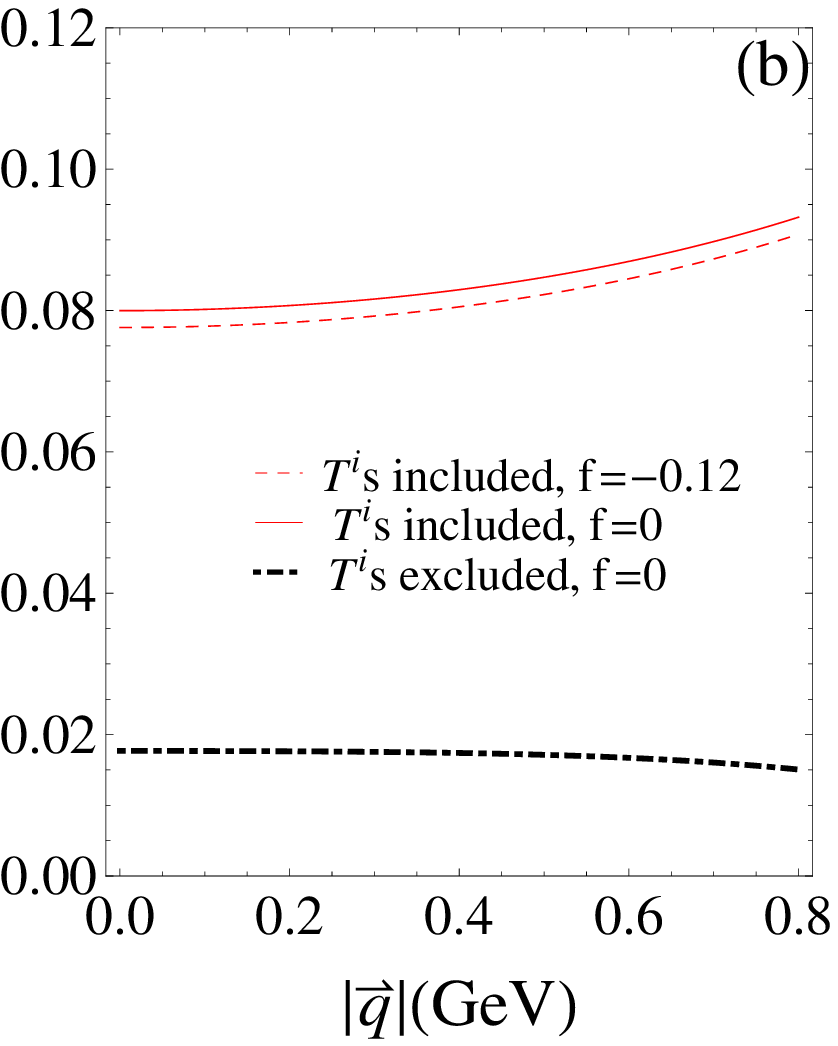}
\caption{(Color online) (a) Borel mass and (b) $\vert \vec{q} \vert$
dependence of $E^{\textrm{sym}}_{V,\rho}$. The unit of the vertical
axis is GeV.}\label{nse}
\end{figure}

$E^{\textrm{sym}}_{V,\rho}$ is plotted in Fig.~\ref{nse}. One finds
that $E^{\textrm{sym}}_{V,\rho}$ ranges from 15 to 80 MeV, which
agrees with previous studies in order of magnitude. The results in
the figure also show that including the twist-4 contribution
slightly enhances the nuclear symmetry energy.

In Fig.~\ref{nse}(b), one finds that $E^{\textrm{sym}}_{V,\rho}$ do
not depend strongly on the quasi nucleon three-momentum up to
$0.5~\textrm{GeV}$. This result agrees with the quasi nucleon
three-momentum dependence of the quasi nucleon self-energy. When the
second set of Table~\ref{table2} is used for the $T^i$s, we find
that $E^{\textrm{sym}}_{V,\rho}$ depends strongly on the quasi
nucleon three-momentum compared to the case when the first set is
used.

\begin{figure}
\includegraphics[width=4.23cm,height=5.1cm]{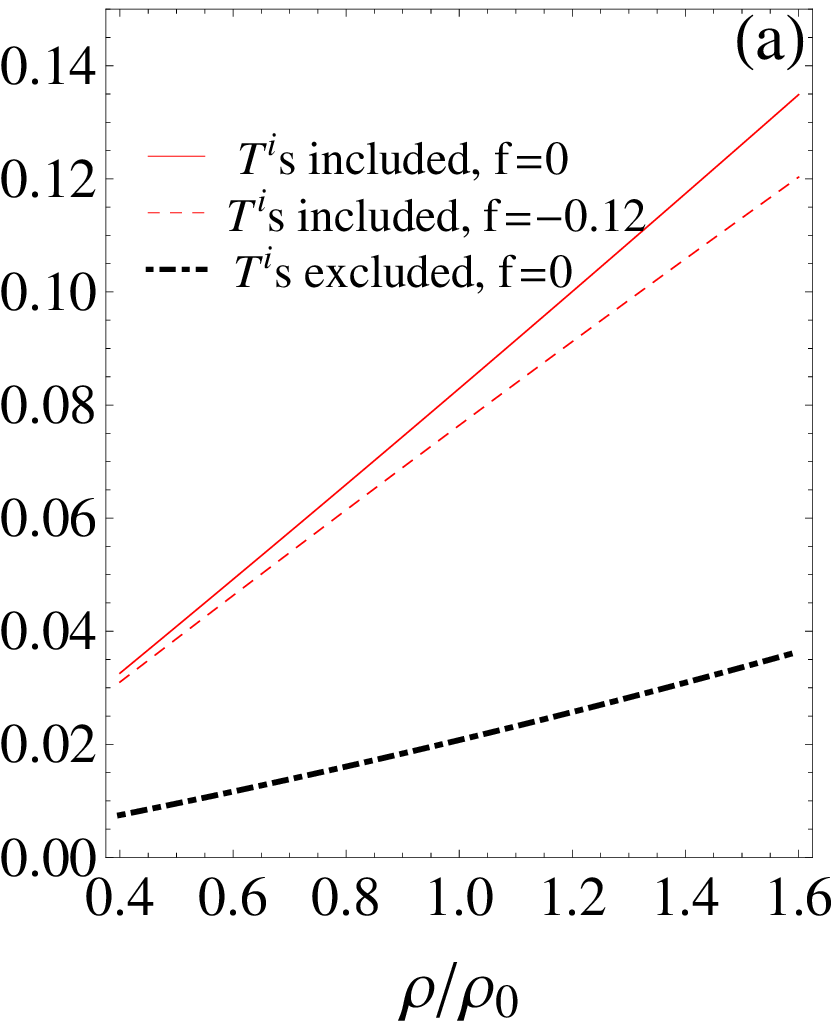}
\includegraphics[width=4.31cm,height=5.25cm]{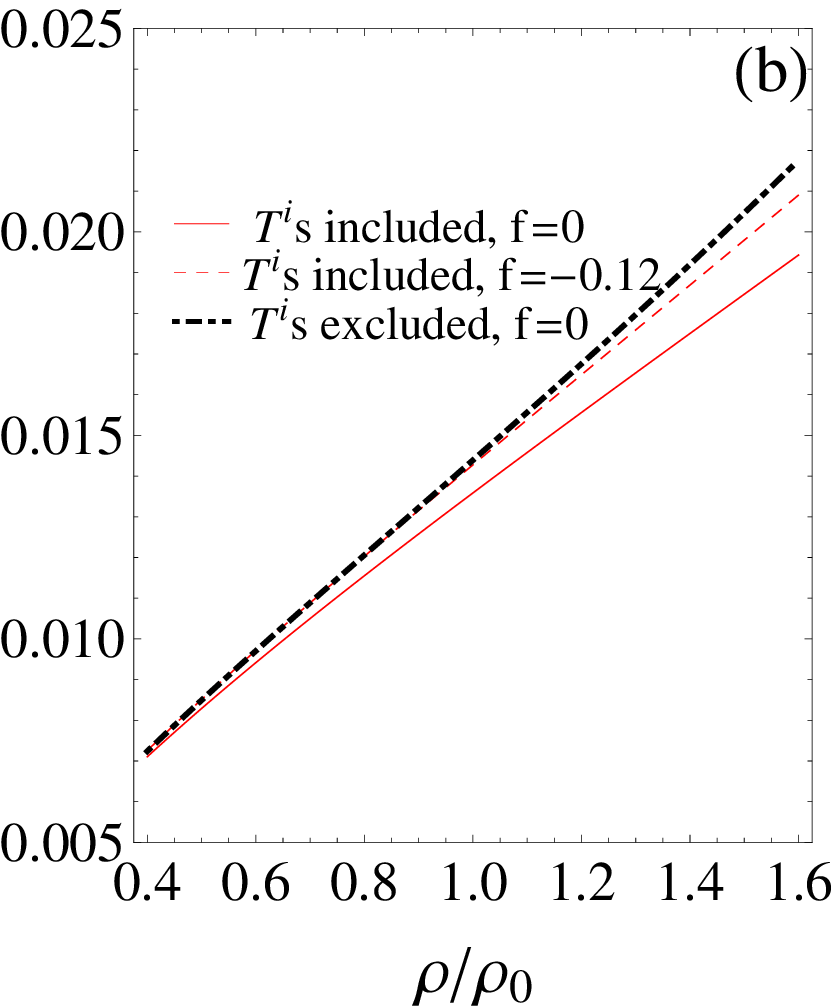}
\caption{(Color online) Density dependence of (a)
$E^{\textrm{sym}}_{V,\rho^2}$ and (b) $E^{\textrm{sym}}_K$. The unit
of the vertical axis is GeV.}\label{nsed}
\end{figure}

One can also work out $E^{\textrm{sym}}_{V,\rho^2}$, although with
larger uncertainty than that for the $E^{\textrm{sym}}_{V,\rho}$.
The density dependence of $E^{\textrm{sym}}_{V,\rho^2}$ and
$E_K^{\textrm{sym}}$ for $0.4\leq\rho/\rho_0\leq1.6$ are plotted in
Fig.~\ref{nsed}.  Here again, the four-quark condensates contribute
nontrivially to the density behavior of $E_V^{\textrm{sym}}$. For
$f=0$, the contribution of $T^i$s gives strong enhancement to
$E^{\textrm{sym}}_{V,\rho^2}$ at higher nuclear density while for
$f=-0.12$, it gives reduction to $E^{\textrm{sym}}_{V,\rho^2}$ at
higher density. This means that the scalar four-quark operators
contribute importantly in providing attraction to the nuclear
symmetry energy. However, $E_K^{\textrm{sym}}$ is slightly reduced
by $T^i$s as the twist-4 matrix elements enhance $M^{*}_N/M_N$. The
parameter set with $f<0$ contributes differently to
$E_V^{\textrm{sym}}$ and $E_K^{\textrm{sym}}$; reduction for
$E_V^{\textrm{sym}}$ and enhancement for $E_K^{\textrm{sym}}$ for
$0.4\leq\rho/\rho_0\leq1.6$.

\begin{figure}[b]
\includegraphics[width=4.27cm,height=5.125cm]{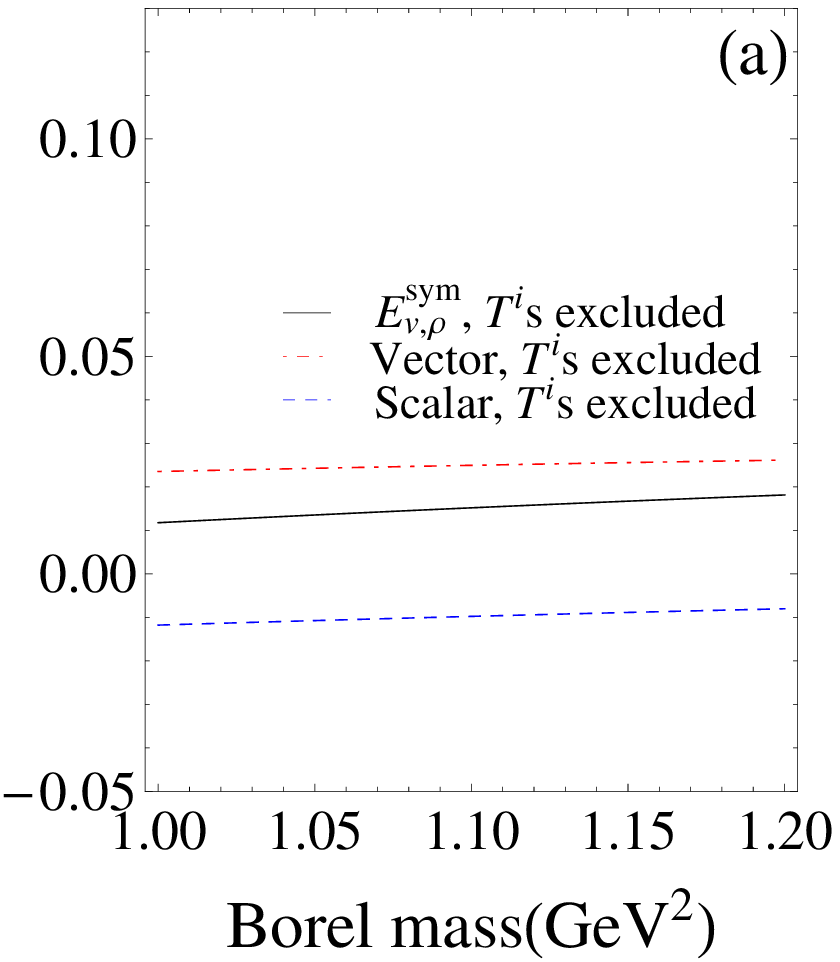}
\includegraphics[width=4.27cm,height=5.125cm]{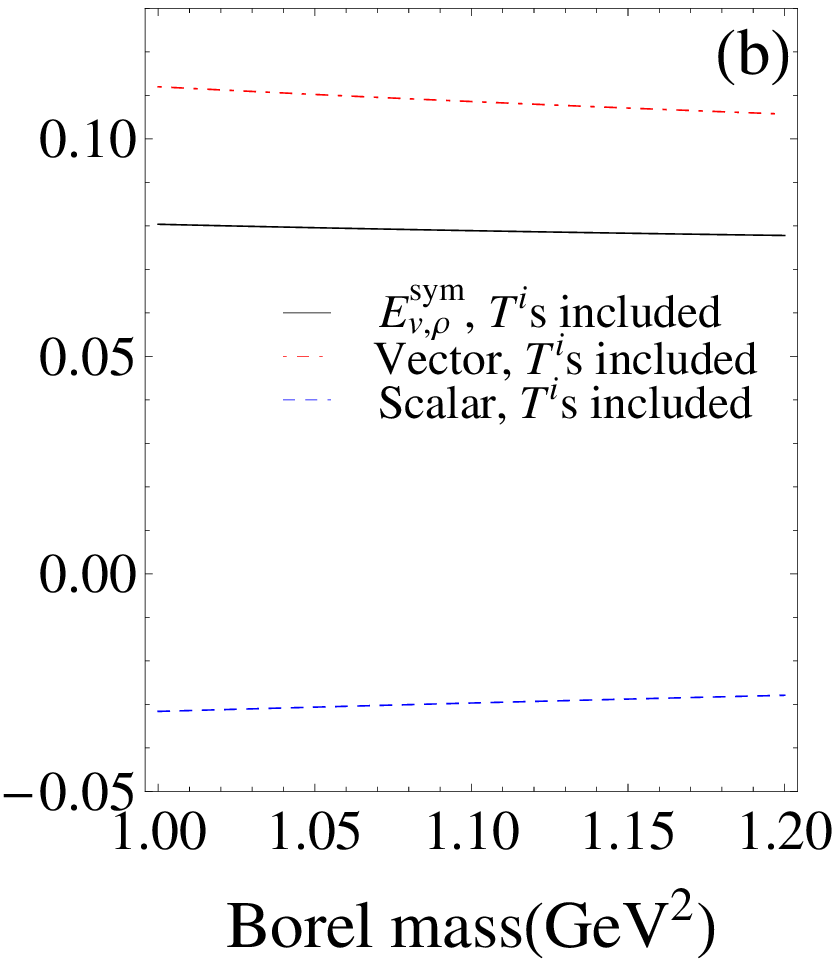}
\caption{(Color online) Scalar-vector self-energy decomposition of
$E^{\textrm{sym}}_{V,\rho}$ (a) without twist-4 contribution and (b)
with twist-4 contribution. The unit of the vertical axis is
GeV.}\label{nsedcmp}
\end{figure}

In Fig.~\ref{nsedcmp}, we plot the scalar
$\left(\frac{\bar{\mathcal{B}}[\Pi_s(q_0^2,\vert\vec{q}\vert)]}{\bar{\mathcal{B}}[\Pi_q(q_0^2,\vert\vec{q}\vert)]}\right)$
and vector
$\left(\frac{\bar{\mathcal{B}}[\Pi_u(q_0^2,\vert\vec{q}\vert)]}{\bar{\mathcal{B}}[\Pi_q(q_0^2,\vert\vec{q}\vert)]}\right)$
self-energy part of $E_V^{\textrm{sym}}$.  In Fig.~\ref{nsedcmp}(a),
we plot the result without the twist-4 contribution, while in
Fig.~\ref{nsedcmp}(b), we include the contribution from twist-4
matrix elements. While both the scalar and vector self-energy give
weak contribution to the self-energy in Fig.~\ref{nsedcmp}(a), one
finds that in Fig.~\ref{nsedcmp}(b), the scalar and vector give
enhanced negative and positive contributions, respectively. The
result shown in Fig.~\ref{nsedcmp}(b) is consistent with the general
trends in RMFT results~\cite{Baran:2004ih}, which show that the
scalar self energy part gives a negative contribution and the vector
self-energy part gives a positive contribution from the exchange of
$\delta$ and $\rho$ meson exchanges, respectively.  One can infer
from this result that the twist-4 contribution mimics the exchange
of the $\delta$ and $\rho$ meson and that it constitutes an
essential part in the origin of the nuclear symmetry energy from
QCD.

\subsubsection{Uncertainties}

\begin{figure}
\includegraphics[width=4.27cm,height=5.125cm]{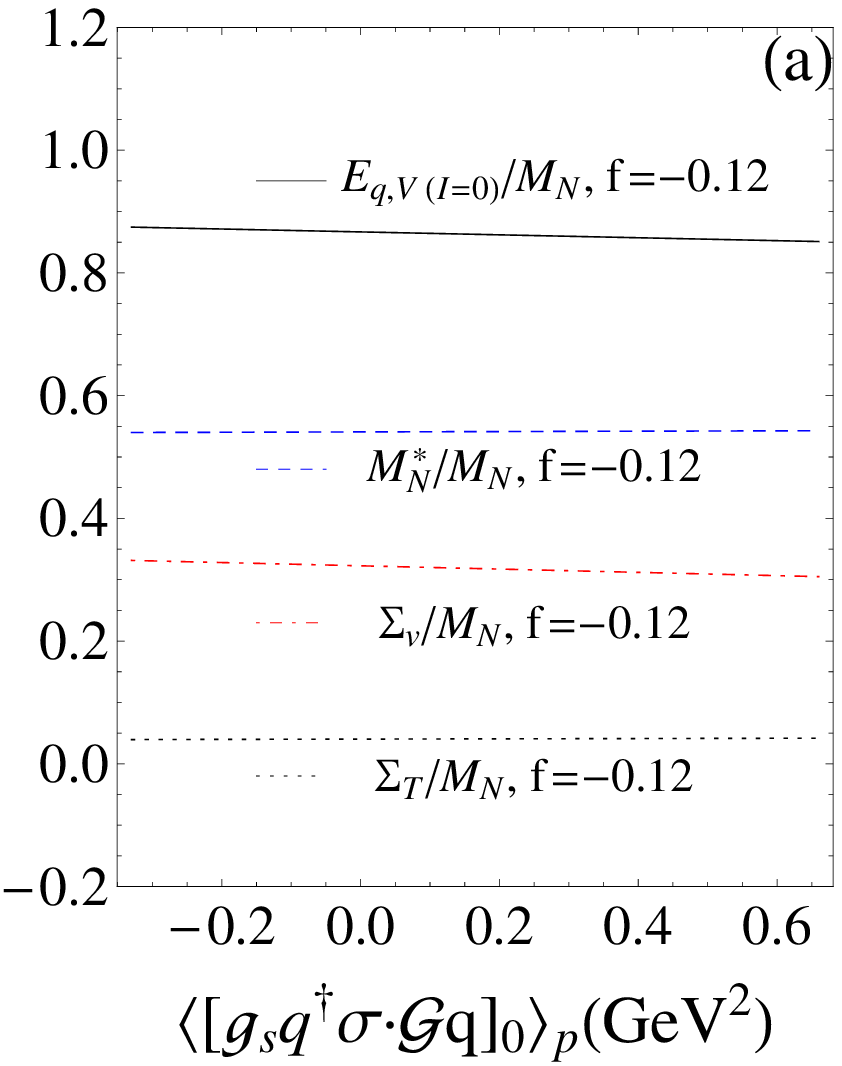}
\includegraphics[width=4.27cm,height=5.125cm]{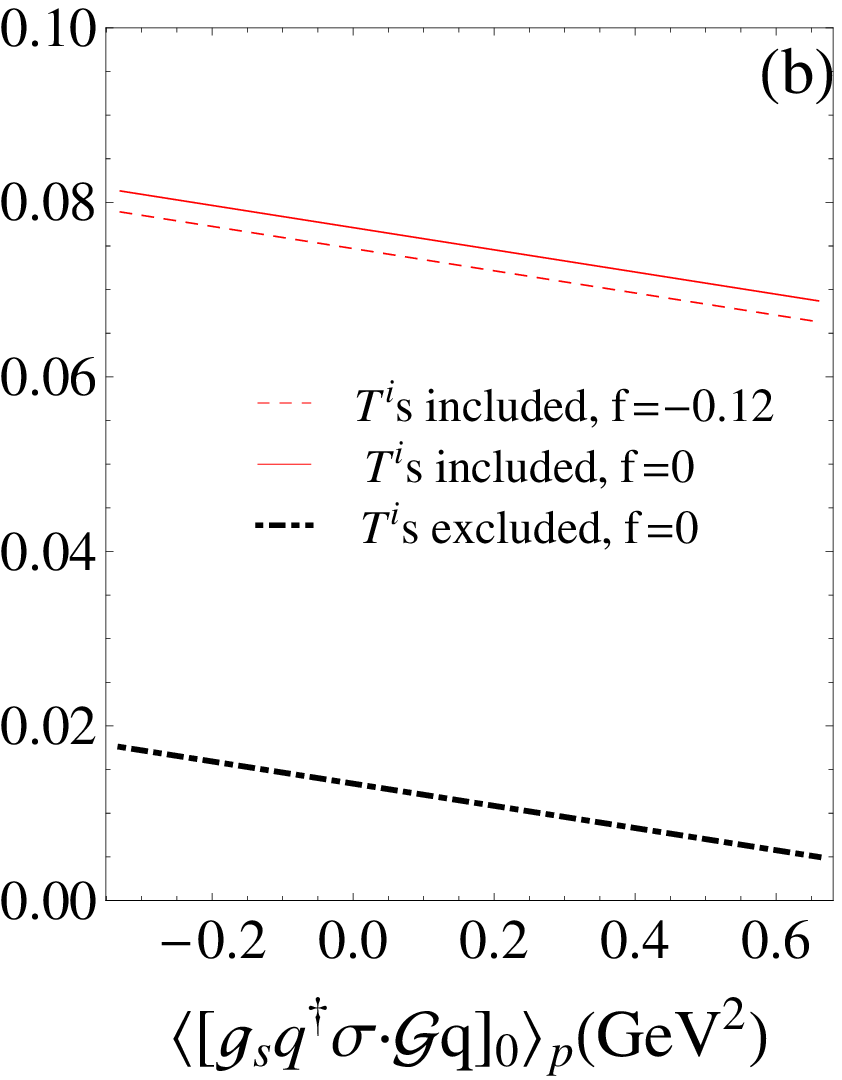}
\caption{(Color online) Sensitivity analysis under variation of the
matrix element $\langle[g_s q^\dagger\sigma\cdot\mathcal{G}q]_0
\rangle_p$  on  (a) $E_{q,V(I=0)}$ and on (b)
$E^{\textrm{sym}}_{V,\rho}$. The unit of the vertical axis for the
right figure is GeV.}\label{nsrl}
\end{figure}

In general, there are two quark-gluon mixed operators with
contracted spin indices, $\langle[g_s
\bar{q}\sigma\cdot\mathcal{G}q]_0 \rangle_p$ and $\langle[g_s
q^\dagger\sigma\cdot\mathcal{G}q]_0 \rangle_p$, which are not
accurately determined. $\langle[g_s
\bar{q}\sigma\cdot\mathcal{G}q]_0 \rangle_p$ does not appear in our
sum rule with Ioffe's nucleon interpolating current
[Eq.~\eqref{crnt}]. As for the operator $\langle[g_s
q^\dagger\sigma\cdot\mathcal{G}q]_0 \rangle_p$, the proton
expectation value has been estimated in
Refs.~\cite{Jin:1992id,Jin:1993up,Shuryak:1981kj,Braun:1986ty} to be
in the range of $-0.33~\textrm{GeV}^2\leq \langle[g_s
q^\dagger\sigma\cdot\mathcal{G}q]_0
 \rangle_p\leq0.66~\textrm{GeV}^2$. Hence, we investigate the $\langle[g_s q^\dagger\sigma\cdot\mathcal{G}q]_0
\rangle_p$ dependence in Fig.~\ref{nsrl}. The matrix element
$\langle[g_s q^\dagger\sigma\cdot\mathcal{G}q]_0 \rangle_p$ does not
give an important contribution to the quasi nucleon self-energies in
the range $-0.33~\textrm{GeV}^2\leq \langle[g_s
q^\dagger\sigma\cdot\mathcal{G}q]_0
\rangle_p\leq0.66~\textrm{GeV}^2$ as we are not interested in the
accuracy of 10 MeV. However, such a magnitude in $\langle[g_s
q^\dagger\sigma\cdot\mathcal{G}q]_0 \rangle_p$ gives nontrivial
fractional change to $E^{\textrm{sym}}_{V,\rho}$. We choose the
value as $\langle[g_s q^\dagger\sigma\cdot\mathcal{G}q]_0
\rangle_p=-0.33~\textrm{GeV}^2$ in this study as was done in
Refs.~\cite{Jin:1992id,Jin:1993up}.

\begin{figure}
\includegraphics[width=4.27cm,height=5.125cm]{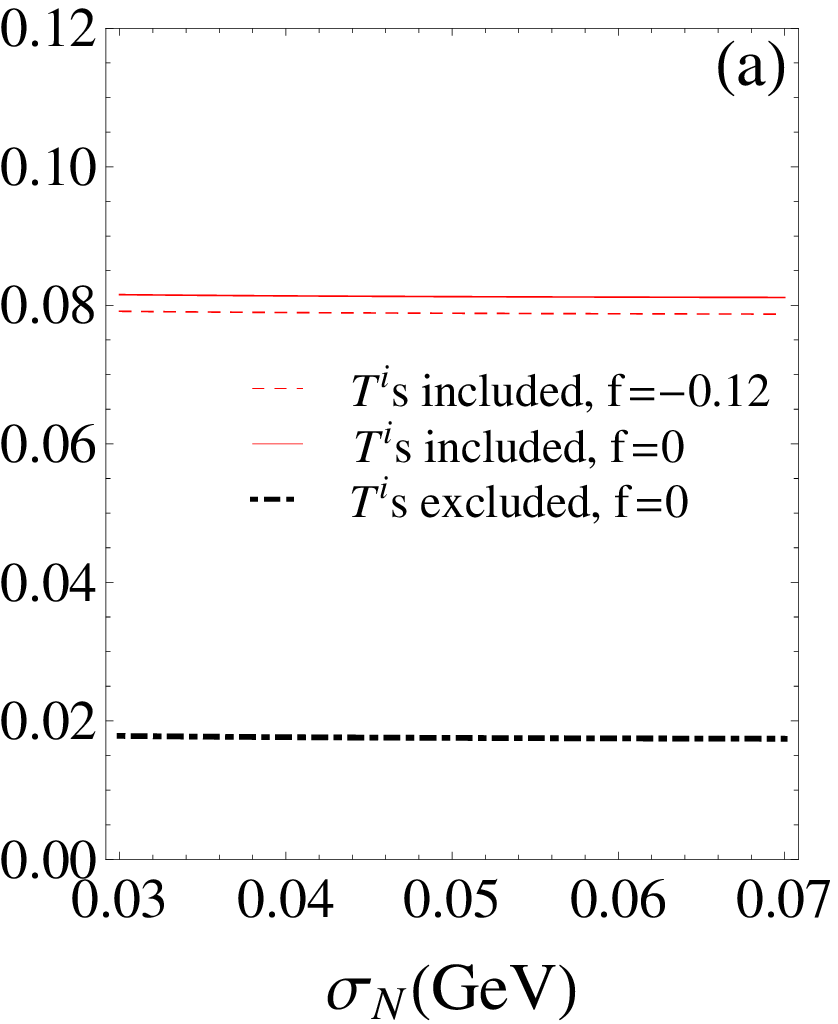}
\includegraphics[width=4.27cm,height=5.125cm]{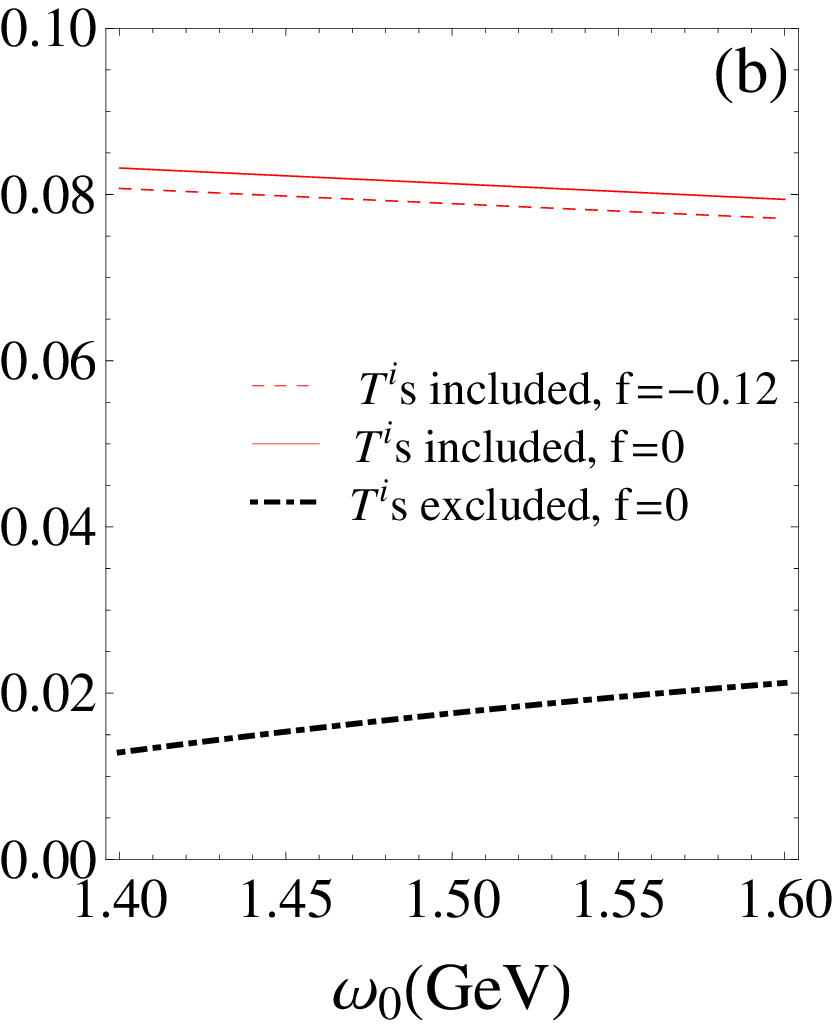}
\caption{(Color online) Variation of (a) $E_{q,V(I=0)}$ under change
in  $\sigma_N$ and (b) $E^{\textrm{sym}}_{V,\rho}$ under change in
$\omega_0$. The unit of the vertical axis is GeV.}\label{sent}
\end{figure}
In our analysis, we fixed the $\sigma$ term to be
$\sigma_N=45~\textrm{MeV}$. In Fig.~\ref{sent}(a) we show that
changing this number from $30~\textrm{MeV} \leq\sigma_N\leq
70~\textrm{MeV}$ changes  $E^{\textrm{sym}}_{V,\rho}$ by less than
5\%.  Also, in principle, the density dependence of the operators
could also induce changes in the continuum.  To investigate this
possibility, we have allowed the continuum to vary
$1.4~\textrm{GeV}\leq\omega_0\leq1.6~\textrm{GeV}$.  As can be seen
in the three lines in Fig.~\ref{sent}(b), the change in the symmetry
energy is less than 10\%.  This suggest that reasonable density
dependence will not appreciably modify the current result.

\subsubsection{Comparison with the result from PCQM}

\begin{figure}[b]
\includegraphics[width=4.25cm,height=5.27cm]{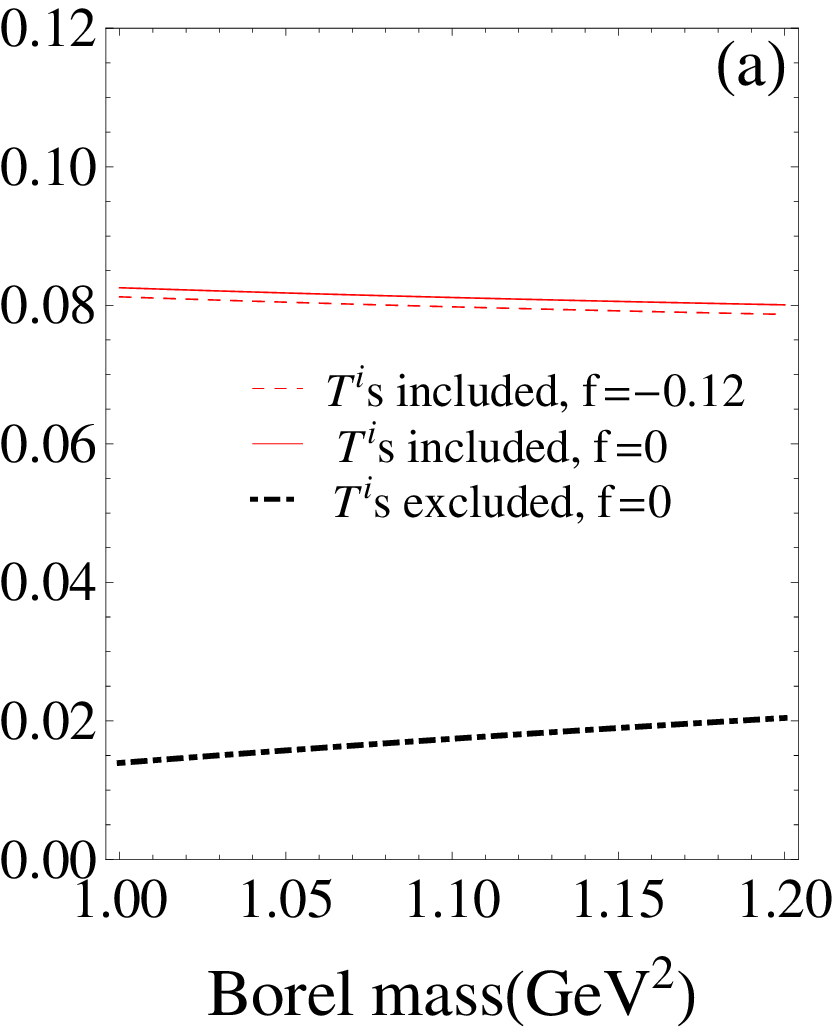}
\includegraphics[width=4.29cm,height=5.15cm]{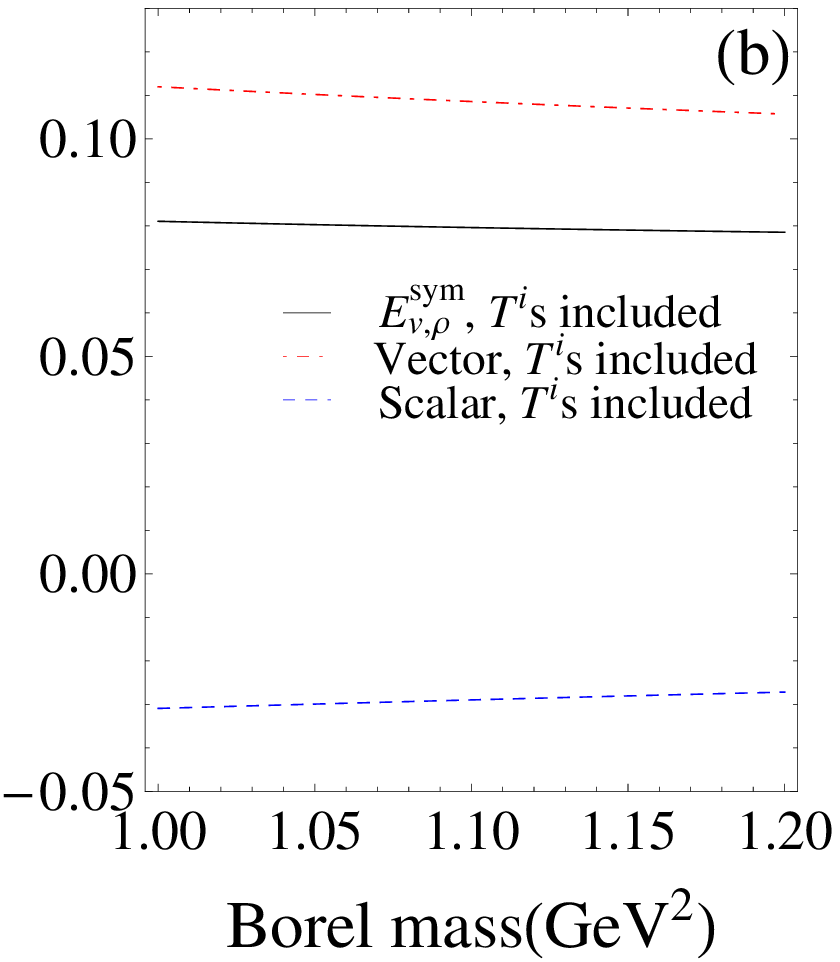}
\caption{(Color online) $E^{\textrm{sym}}_{V,\rho}$ in which
$\langle[\bar{q}q]_1\rangle_p$ is replaced with $\frac{1}{2}\zeta_N$
from PCQM~\cite{Lyubovitskij:2000sf}. (a)
$E^{\textrm{sym}}_{V,\rho}$ with $\zeta_N=0.54$ and  (b)
Scalar-vector decomposition of $E^{\textrm{sym}}_{V,\rho}$ with
$\zeta_N=0.54$ (including $T^i$s). The unit of the vertical axis is
GeV.}\label{pcqm}
\end{figure}
As mentioned in the Introduction, there were early studies about
nucleon sum rule in the asymmetric nuclear matter using the first
approach \cite{Drukarev:2004fn,Drukarev:2010xv,Drukarev:2012av}. In
comparison with this study, the two main differences are the
followings. First, in
Refs.~\cite{Drukarev:2004fn,Drukarev:2010xv,Drukarev:2012av},the OPE
expansion was performed in the light cone direction and the $q^2$
dispersion relation was used.  On the other hand, in this work, the
OPE is a short distance expansion and the energy dispersion relation
is used; consequently, the OPE totally differs. Second, in
Refs.~\cite{Drukarev:2004fn,Drukarev:2010xv,Drukarev:2012av}
$\langle[\bar{q}q]_1\rangle_p$ is obtained from PCQM
\cite{Lyubovitskij:2000sf,Drukarev:2004fn} while we calculate
$\langle[\bar{q}q]_1\rangle_p$ from the leading  chiral
expansion~\cite{Shifman:1978zn,Choi:1991}. Because the OPEs totally
differ, it is impossible to compare both results term by term in
terms of the QCD condensates, but here we can compare the final
results in $E^{\textrm{sym}}_{V,\rho}$. From a phenomenological
aspect, Refs.~\cite{Drukarev:2004fn,Drukarev:2010xv} give values for
the nuclear symmetry energy,
$E^{\textrm{sym}}_{V}+E^{\textrm{sym}}_{K}=29~\textrm{MeV}$, which
almost agrees with the phenomenological estimates. As one can check
in Fig.~\ref{pcqm}, using the same values for
$\langle[\bar{q}q]_1\rangle_p=\frac{1}{2}\zeta_N$ as estimated from
the PCQM~\cite{Lyubovitskij:2000sf}, we find
$E^{\textrm{sym}}_{V,\rho} \sim 80~\textrm{MeV}$, there is no
significant change. However, in our approach, we find interesting
similarities with the main results from RMFT~\cite{Baran:2004ih},
namely strong vector repulsion and scalar attraction.

\section{Conclusion}

In this paper we studied the nuclear bulk properties in asymmetric
nuclear matter by calculating the quasi nucleon self-energies with
QCD sum rule approach. In particular, we identified all the twist-4
local condensates appearing in the nucleon sum rule. Using the
existing estimates for the twist-4 matrix element from DIS, we were
able to find the magnitudes of all the twist-4 matrix elements
($T^i$) in our sum rule except for two mixed-quark-flavor-type
condensates.  We have calculated the nuclear symmetry energy and
found that twist-4 contributions  are non-negligible and essential
to give a phenomenologically consistent result with RMFT for  the
quasi nucleon self-energy and the nuclear symmetry energy.

For the symmetric nuclear matter case, we  found that $E_{q,V(I=0)}$
is enhanced by $\sim 50~\textrm{MeV}$ with $T^i$s in the first set
of Table~\ref{table2}.  Because the $T^i$s in the  first set of
Table~\ref{table2} provides qualitatively reliable sum rule results
while $T^i$s in the second set of Table~\ref{table2} do not, we
conclude that taking the sum rule results with $T^i$s in the first
set is the reasonable choice. With parameter set $f<0$,
dimension-six spin-0 (scalar) operators reduces $E_{q,V(I=0)}/M_N$
to $\sim0.87$.

For the asymmetric nuclear matter case, we confirmed two meaningful
facts. First, the QCD sum rule technique can be used to successfully
reproduce the acceptable result for the nuclear symmetry energy at
the nuclear matter density. Second, dimension-six spin-2 (twist-4)
condensates play important roles in making the scalar part
contribute negatively to the self-energy and, thus, providing a
consistent picture  for the $E^{\textrm{sym}}$ with the RMFT
results~\cite{Baran:2004ih},
\begin{align}
E^{\textrm{sym}}_{V}=\frac{1}{2}\left[f_\rho-f_\delta\left(
\frac{m^{*}}{E^{*}_F}\right)\right]\rho_B,
\end{align}
where $f_\rho$ is the isovector $\rho$ meson coupling, $f_\delta$
the isoscalar $\delta$ ($f_0$) coupling and $\rho_B$ the nuclear
matter density. Moreover, our approach provides a first attempt to
understanding the origin of $E^{\textrm{sym}}_{V,\rho}$ in terms of
local operators directly from QCD. This extends the analogy between
QCD sum rules to RMFT for the symmetric nuclear matter established
in Refs.~\cite{Cohen:1991js,Jin:1992id,Jin:1993up} to the asymmetric
limit.

While the uncertainties in $T^i$s and in the four-quark scalar operators with
the $f$ parametrization are still large, attempts to measure the twist-4
contribution in DIS at the future upgrade at Jefferson Lab is
expected to lower the uncertainties and provide more insights to the
value for the nuclear expectation value of the four-quark operators.

%\begin{acknowledgements}
\section*{ACKNOWLEDGEMENTS}
This work was supported by Korea national research foundation under
Grants No. KRF-2011-0030621 and No. KRF-2011-0020333.  We also thank
S. Choi for providing materials from his masters thesis.

%\end{acknowledgements}

\appendix

\section{Baryon octet mass relation}
In this section, we summarize an essential argument for obtaining
$\langle[\bar{q}q]_1\rangle_p=\frac{1}{2}\left(\langle
p\vert\bar{u}u\vert p\rangle-\langle p\vert\bar{d}d\vert
p\rangle\right)$ from Ref.~\cite{Choi:1991} and
Ref.~\cite{Shifman:1978zn}. Phenomenologically, the nucleon mass can
be expressed in terms of the matrix element of the trace of the
energy-momentum tensor:
\begin{align}
m_N\langle N\vert \bar{\psi}_N \psi_N\vert N \rangle=&\langle
N\vert\theta^\mu_{~\mu}\vert N \rangle.
\end{align}
Using the equations of motion, the trace of the energy-momentum
tensor can be written as
\begin{align}
\theta^\mu_{~\mu}=&~m_u\bar{u}u+m_d\bar{d}d+m_s\bar{s}s+\sum_{h=c,t,b}
m_h\bar{h}h+\cdots\nonumber\\
=&\left(\frac{\bar{\beta}}{4\alpha_s}\right)G^2+m_u\bar{u}u+m_d\bar{d}d+m_s\bar{s}s+O(\mu^2/4m_h^2),\label{trace}
\end{align}
where the $h$'s are the heavy quark fields, and the gluonic term comes from
the trace anomaly~\cite{Collins:1976yq, Crewther:1972kn,
Chanowitz:1972da}. $\bar{\beta}=-9\alpha_s^2/2\pi$ is the
``reduced'' Gellmann-Low function in which heavy quark contribution
has been subtracted out using the heavy quark
expansion~\cite{Witten:1975bh}.

Eqation~\eqref{trace} can be applied to the lowest-lying baryon
octet. The baryon octet mass relations to first order in SU(3)
flavor symmetry breaking are as follows:
\begin{align}
m_p&=A+m_uB_u+m_dB_d+m_sB_s,\nonumber\\
m_n&=A+m_uB_d+m_dB_u+m_sB_s,\nonumber\\
m_{\Sigma^{+}}&=A+m_uB_u+m_dB_s+m_sB_d,\nonumber\\
m_{\Sigma^{-}}&=A+m_uB_s+m_dB_u+m_sB_d,\nonumber\\
m_{\Xi^{0}}&=A+m_uB_d+m_dB_s+m_sB_u,\nonumber\\
m_{\Xi^{-}}&=A+m_uB_s+m_dB_d+m_sB_u,\label{baryonoctet}
\end{align}
where $A\equiv\langle(\bar{\beta}/4\alpha_s)G^2\rangle_p$,
$B_u\equiv\langle\bar{u}u\rangle_p$,
$B_d\equiv\langle\bar{d}d\rangle_p$ and
$B_s\equiv\langle\bar{s}s\rangle_p$. In this relation, correction
terms for hyperon is neglected~\cite{Cheng:1990}.
From~\eqref{baryonoctet} one can obtain
\begin{align}
\langle p\vert\bar{u}u\vert p\rangle-\langle p\vert\bar{d}d\vert
p\rangle=\frac{(m_{\Xi^{0}}+m_{\Xi^{-}})-(m_{\Sigma^{+}}+m_{\Sigma^{-}})}{2m_s-(m_u+m_d)}.\label{ratio1}
\end{align}

\section{A simple constraint for twist-4 operators from zero identity}

In this section, we show an explicit calculation for a simple
constraint using the zero identity~\cite{Thomas:2007gx}. For the single
quark flavor diquark structure,
\begin{align}
\epsilon_{abc}(u^T_a C\Gamma u_b)=0,\qquad\textrm{if
$(C\Gamma)^T=-C\Gamma$},\label{zeroidnty}
\end{align}
where ($\Gamma=\{I,\gamma_5,i\gamma_\mu \gamma_5\}$) satisfies the above
condition. Therefore, constraints for the four-quark operator can be
obtained by requiring that the Fierz transformed form of the
products of above diquarks are zero.  An example is the following:
\begin{align}
\epsilon_{abc}&\epsilon_{a'b'c}(u^T_a C\gamma_\mu\gamma_5
u_b)(\bar{u}_{b'} \gamma_\nu \gamma_5 C \bar{u}^T_{a'})\nonumber\\
=&~\epsilon_{abc}\epsilon_{a'b'c}~\frac{1}{16}~(\bar{u}_{a'}
\Gamma^o u_{a})(\bar{u}_{b'} \Gamma^k
u_{b})\nonumber\\
&\times\textrm{Tr}\left[
\gamma_\mu \gamma_5 \Gamma_k \gamma_\nu \gamma_5 C \Gamma_o^T C\right]\nonumber\\
=&~\epsilon_{abc}\epsilon_{a'b'c}~\frac{1}{16}~\bigg\{(\bar{u}_{a'}u_a)(\bar{u}_{b'}u_b)(4g_{\mu\nu})\nonumber\\
&+
(\bar{u}_{a'}\gamma_5u_a)(\bar{u}_{b'}\gamma_5u_b)(-4g_{\mu\nu})\nonumber\\
& +(\bar{u}_{a'}\gamma^{\alpha}u_a)(\bar{u}_{b'}\gamma^{\beta}
u_b)(4S_{\mu\beta\nu\alpha})\nonumber\\
&-(\bar{u}_{a'}\gamma^{\alpha}\gamma_5u_a)(\bar{u}_{b'}\gamma^{\beta} \gamma_5 u_b)(4S_{\mu\beta\nu\alpha})\nonumber\\
&-(\bar{u}_{a'}\sigma^{\alpha\bar{\alpha}}u_a)
(\bar{u}_{b'}\sigma^{\beta\bar{\beta}}u_b)\frac{1}{4}\textrm{Tr}\left[
\gamma_\mu \sigma_{\beta\bar{\beta}} \gamma_\nu
\sigma_{\alpha\bar{\alpha}}\right]\nonumber\\
&+(\bar{u}_{a'}\gamma^{\alpha}u_a)
(\bar{u}_{b'}\gamma^{\beta}\gamma_5
u_b)(8i\epsilon_{\mu\beta\nu\alpha})\nonumber\\
&+(\bar{u}_{a'}u_a) (\bar{u}_{b'}\sigma^{\alpha\bar{\alpha}}u_b)(8i
g_{\alpha\mu}g_{\bar{\alpha}\nu})\nonumber\\
&+(\bar{u}_{a'}\gamma_5u_a)
(\bar{u}_{b'}\sigma^{\alpha\bar{\alpha}}u_b)
(4\epsilon_{\mu\nu\alpha\bar{\alpha}})\bigg\}=0,\label{constraint1}
\end{align}
where $S_{\mu\alpha\nu\beta}=g_{\mu\alpha}g_{\nu\beta}+
g_{\mu\beta}g_{\alpha\nu}-g_{\mu\nu}g_{\alpha\beta}$. By subtracting
Eq.~\eqref{constraint1} from Eq.~\eqref{ope-4q-1},
Eq.~\eqref{ope-4q-1} can be simplified into Eq.~\eqref{4qope1}.

\section{Estimation of twist-4 matrix elements}
In this section, we provide a detailed treatment for extracting
$T^i$s from the values estimated in Ref.~\cite{Choi:1993cu}. In
Ref.~\cite{Choi:1993cu}, twist-4 operators which appear in our
nucleon sum rule are given as
\begin{align}
\frac{1}{4\pi\alpha_s}&\frac{M_N}{2}\left(u_\alpha
u_\beta-\frac{1}{4}g_{\alpha \beta}\right)K_u^1\nonumber\\
=&\langle (\bar{u} \gamma_\alpha \gamma_5 t^A u)(\bar{u}
\gamma_\beta \gamma_5 t^A u)
\rangle_p\vert_{\textrm{s,t}}\nonumber\\
&+ \langle (\bar{u} \gamma_\alpha \gamma_5 t^A u)(\bar{d}
\gamma_\beta \gamma_5 t^A d) \rangle_p\vert_{\textrm{s,t}},\\
\frac{1}{4\pi\alpha_s}&\frac{M_N}{2}\left(u_\alpha
u_\beta-\frac{1}{4}g_{\alpha \beta}\right)K_u^2\nonumber\\
=&\langle (\bar{u} \gamma_\alpha  t^A u)(\bar{u} \gamma_\beta t^A u)
\rangle_p\vert_{\textrm{s,t}} \nonumber\\
&+ \langle (\bar{u} \gamma_\alpha t^A u)(\bar{d} \gamma_\beta  t^A
d)
\rangle_p\vert_{\textrm{s,t}},\\
\frac{1}{4\pi\alpha_s}&\frac{M_N}{2}\left(u_\alpha
u_\beta-\frac{1}{4}g_{\alpha \beta}\right)K_{ud}^1
\nonumber\\
=&2\langle (\bar{u} \gamma_\alpha \gamma_5 t^A u)(\bar{d}
\gamma_\beta \gamma_5 t^A d) \rangle_p\vert_{\textrm{s,t}},
\end{align}
where we changed the normalization for the nucleon state appearing
in Ref.~\cite{Choi:1993cu} to the following:
\begin{align}
\langle N (p) \vert
N(p')\rangle=\frac{\omega_p}{M_N}(2\pi)^3\delta^3(\vec{p}-\vec{p'}),
\end{align}
with $\omega_p=p_0=\sqrt{\vec{p}^2+M^2_N}$. Here only $K_{ud}^1$ is
uniquely determined: $K_{ud}^1=-0.083~\textrm{GeV}^2$. One can set a
constraint $\vert K^1_d \vert = \vert K^1_u \vert\cdot\beta < \vert
K_{ud}^1 \vert < \vert K^1_u\vert$ with an ansatz that the ratio
$K^i_d/K^i_u$ is equal to the momentum fraction of the $d$ and $u$
quarks in the nucleon:
\begin{align}
K^i_d/K^i_u\simeq\frac{\int x(d(x)+\bar{d}(x))dx}{\int
x(u(x)+\bar{u}(x))dx}\equiv\beta=0.476.
\end{align}

Varying $K_{u}^1$ with the constraint above, one can estimate
$K_{u}^2$ as a functions of $K_{u}^1$ from the constraints from DIS;
the results are given in  Table~\ref{table3}.

$T^1_{uu}$ and $T^1_{dd}$ can be easily estimated by taking
$T^1_{ud}=\frac{1}{2}K^1_{ud}$
 from $K^1_{u}$ and $K^1_{d}$:
\begin{align}
T^1_{uu}=&~K^1_u-T^1_{ud},\\
T^1_{dd}=&~K^1_d-T^1_{ud},\\
T^1_{ud}=&~\frac{1}{2}K^1_{ud}=-0.042~\textrm{GeV}^2.
\end{align}

Similarly, one can try to obtain $T^2_{uu}$ and $T^2_{dd}$ from
$K^2_{u}$ and $K^2_{d}$. As $T^2_{ud}=\frac{1}{2}K^2_{ud}$ has not
been determined uniquely as $T^1_{ud}=\frac{1}{2}K^1_{ud}$, we
assumed that the ratio $T^1_{uu}/T^1_{dd}$ is equal to
$T^2_{uu}/T^2_{dd}$. Then, by the following relation, one can
estimate $T^2_{qq}$s:
\begin{align}
T^2_{uu}=&~(K^2_u-K^2_d)\left(1-\frac{T^1_{dd}}{T^1_{uu}}\right)^{-1},\\
T^2_{dd}=&\left(\frac{T^1_{dd}}{T^1_{uu}}\right) T^2_{uu},\\
T^2_{ud}=&~\frac{1}{2}\left([K^2_u+K^2_d]-[T^2_{uu}+T^2_{dd}]\right),
\end{align}
where $K^2_u-K^2_d$ and $K^2_u+K^2_d$ can be obtained from
Table~\ref{table3}.

For the single quark flavor case, $T^3_{qq}$ and $T^4_{qq}$ can be
obtained from Eq.~\eqref{cond1} and Eq.~\eqref{cond2}. As discussed
in Ref.~\cite{Jaffe:1981sz}, we neglect $(\bar{u}
\sigma_{o}^{~\alpha} u )(\bar{u} \sigma^{o\beta}
u)\vert_{\textrm{s,t}}$. Then, $T^3_{qq}$ and $T^4_{qq}$ can be
related as
\begin{align}
T^3_{qq}=&-\frac{15}{4}T^1_{qq}+\frac{9}{4}T^2_{qq}\\
T^4_{qq}=&-\frac{15}{4}T^2_{qq}+\frac{9}{4}T^1_{qq}.
\end{align}

$T^i_{qq}$s can be classified as the three different classification of
$K^i$s given in Table~\ref{table3}:
$K^1_u=\{K^1_{ud}/\beta,~K^1_{ud}(\beta+1)/2\beta,~K^1_{ud}\}$ and
$K^1_u=\{-K^1_{ud},~-K^1_{ud}(\beta+1)/2\beta,~-K^1_{ud}/\beta\}$.
$T^i_{qq}$'s are classified in Table~\ref{table2} according to these
three classifications in the two sets.

\begin{widetext}\allowdisplaybreaks

\begin{table}
\begin{tabular}{ l c c c  l c c }
\hline\hline  $K^1_u$ &&$K^2_u$ & &$K^1_u$ && $K^2_u$
\\
\hline $K^1_{ud}/\beta=-0.173 $&&0.203 & &$-K^1_{ud}=0.083$ && -0.181  \\
$K^1_{ud}(\beta+1)/2\beta=-0.112 $&&0.110 & &$-K^1_{ud}(\beta+1)/2\beta=0.112$ && -0.225 \\
$K^1_{ud}=-0.083 $&&0.066 & &$-K^1_{ud}/\beta=0.173 $& & -0.318 \\
\hline \hline
\end{tabular}\caption{Table for $K^i_u$ from Ref.~\cite{Choi:1993cu}. Units are in $\textrm{GeV}^2$.}\label{table3}
\end{table}

\section{QCD sum rule formulas for $E_{q,V(I)}$ and $E^{\textrm{sym}}_{V,\rho}$ }
In this section, we provide the detailed description for
\begin{align}
E_{q,V(I)}&=\frac{\mathcal{N}^{n,p}_{(\rho^0,I^0)} +
\mathcal{N}^{n,p}_{(\rho,I^0)}\rho+
\left[\mathcal{N}^{n,p}_{(\rho,I)}\rho\right]I}{\mathcal{D}^{n,p}_{(\rho^0,I^0)}
+
\mathcal{D}^{n,p}_{(\rho,I^0)}\rho+ \left[\mathcal{D}^{n,p}_{(\rho,I)}\rho\right]I},\\
E^{\textrm{sym}}_{V,\rho}&=\frac{1}{4}\rho
\left[\frac{1}{\mathcal{D}^{p}_{(\rho^0,I^0)}}\left(-2\mathcal{N}^{p}_{(\rho,I)}\right)
-\frac{\mathcal{N}^{p}_{(\rho^0,I^0)}}{(\mathcal{D}^{p}_{(\rho^0,I^0)})^2}
\left(-2\mathcal{D}^{p}_{(\rho,I)}\right)\right],
\end{align}
in QCD sum rule formula. In this formula,
$\mathcal{N}^{p}_{(\rho^m,I^l)}$ and
$\mathcal{D}^{p}_{(\rho^m,I^l)}$ are as follows:

\begin{align}
\mathcal{N}^{p}_{(\rho^0,I^0)}=&-\frac{1}{4\pi^2}(M^2)^2E_1\langle
[\bar{q}q]_0\rangle_{\textrm{vac}},\\
\mathcal{N}^{p}_{(\rho,I^0)}=&-\frac{1}{4\pi^2}(M^2)^2E_1\langle
[\bar{q}q]_0 \rangle_{p}-\frac{4}{3\pi^2}\vec{q}^2\langle [\bar{q}
\{iD_{0} iD_{0}\}
q]_0 \rangle_{p}L^{-\frac{4}{9}}\nonumber\\
&+\frac{2}{3\pi^2}(M^2)^2\langle [q^\dagger
q]_0\rangle_{p}E_1L^{-\frac{4}{9}}+\frac{4}{\pi^2}\vec{q}^2\langle
[\bar{q} \{\gamma_{0} iD_{0} iD_{0}\} q]_0
\rangle_{p}L^{-\frac{4}{9}}-\frac{1}{12\pi^2}M^2\langle[ g_s
q^\dagger\sigma\cdot \mathcal{G}
q]_0\rangle_{p}E_0L^{-\frac{4}{9}}\nonumber\\
&+\bar{E}_q\bigg\{\frac{20}{9\pi^2}M^2\langle [\bar{q} \{\gamma_{0}
iD_{0}\} q]_0 \rangle_{p}E_0L^{-\frac{4}{9}}
-\frac{1}{36\pi^2}M^2\left\langle\frac{\alpha_s}{\pi}[(u\cdot
G)^2+(u\cdot
\tilde{G})^2]\right\rangle_{p}E_0L^{-\frac{4}{9}}\nonumber\\
&+\frac{1}{\pi\alpha_s}\frac{M_N}{2}\left(
 [T^1_{ud}-T^2_{ud}] +
[T^1_{~0}-T^2_{~0}]-\frac{1}{3} [T^3_{~0}-T^4_{~0}]\right)
L^{-\frac{4}{9}}-\frac{4}{3}\langle\bar{q}q\rangle_{\textrm{vac}}\langle[q^\dagger
q]_0\rangle_p\bigg\},\\
\mathcal{N}^{p}_{(\rho,I)}=&-\frac{1}{4\pi^2}(M^2)^2E_1\langle
[\bar{q}q]_1 \rangle_{p}-\frac{4}{3\pi^2}\vec{q}^2\langle [\bar{q}
\{iD_{0} iD_{0}\}
q]_1 \rangle_{p}L^{-\frac{4}{9}}\nonumber\\
&-\frac{1}{2\pi^2}(M^2)^2\langle [q^\dagger
q]_1\rangle_{p}E_1L^{-\frac{4}{9}}-\frac{2}{\pi^2}\vec{q}^2\langle
[\bar{q} \{\gamma_{0} iD_{0} iD_{0}\} q]_1
\rangle_{p}L^{-\frac{4}{9}}+\frac{1}{4\pi^2}M^2\langle [g_s
q^\dagger\sigma\cdot \mathcal{G} q]_1
\rangle_{p}E_0L^{-\frac{4}{9}}\nonumber\\
&+\bar{E}_q\left\{\frac{4}{3\pi^2}M^2\langle [\bar{q} \{\gamma_{0}
iD_{0}\} q]_1
\rangle_{p}E_0L^{-\frac{4}{9}}+\frac{1}{\pi\alpha_s}\frac{M_N}{2}\left(
-[T^1_{~1}-T^2_{~1}]+\frac{1}{3}
[T^3_{~1}-T^4_{~1}]\right)+\frac{4}{3}\langle\bar{q}q\rangle_{\textrm{vac}}\langle[q^\dagger
q]_1\rangle_p\right\}
L^{-\frac{4}{9}},\\
\mathcal{D}^{p}_{(\rho^0,I^0)}=&\frac{1}{32\pi^4}(M^2)^3E_2L^{-\frac{4}{9}}
+\frac{1}{32\pi^2}M^2\left\langle\frac{\alpha_s}{\pi}G^2\right\rangle_{\textrm{vac}}E_0L^{-\frac{4}{9}}
+\frac{2}{3}\langle[\bar{q}q]_0\rangle^2_{\textrm{vac}} L^{\frac{4}{9}},\\
\mathcal{D}^{p}_{(\rho,I^0)}=&-\left(\frac{5}{9\pi^2}M^2E_0-\frac{8}{9\pi^2}\vec{q}^2\right)\langle[
\bar{q} \{\gamma_{0} iD_{0}\} q]_0
\rangle_{p}L^{-\frac{4}{9}}\nonumber\\
&+\frac{1}{32\pi^2}M^2\left\langle\frac{\alpha_s}{\pi}G^2\right\rangle_{p}E_0L^{-\frac{4}{9}}+\frac{1}{144\pi^2}\left(M^2E_0
-4\vec{q}^2\right)\left\langle\frac{\alpha_s}{\pi}[(u\cdot
G)^2+(u\cdot \tilde{G})^2]\right\rangle_{p}L^{-\frac{4}{9}}\nonumber\\
&+
\frac{4}{3}f\langle\bar{q}q\rangle_{\textrm{vac}}\langle[\bar{q}q]_0
\rangle_pL^{\frac{4}{9}} -\frac{1}{4\pi\alpha_s}\frac{M_N}{2}\left(
 [T^1_{ud}-T^2_{ud}]
+ [T^1_{~0}-T^2_{~0}]-\frac{1}{3} [T^3_{~0}-T^4_{~0}]\right)
L^{-\frac{4}{9}}\nonumber\\
&+\bar{E}_q\bigg\{\frac{1}{3\pi^2}M^2E_0L^{-\frac{4}{9}}\langle
[q^\dagger
q]_0\rangle_{p}-\frac{4}{3\pi^2}\left(1-\frac{\vec{q}^2}{M^2}\right)\langle
[\bar{q} \{\gamma_{0} iD_{0} iD_{0}\} q]_0
\rangle_{\rho,I}L^{-\frac{4}{9}}\nonumber\\
&-\frac{2}{3\pi^2}\langle [\bar{q} \{\gamma_{0} iD_{0} iD_{0}\}
q]_0u \rangle_{p}L^{-\frac{4}{9}}+\frac{1}{18\pi^2}\langle [g_s
q^\dagger\sigma\cdot \mathcal{G}
q]_0\rangle_{p}L^{-\frac{4}{9}}\bigg\},\\
\mathcal{D}^{p}_{(\rho,I)}=&\frac{1}{3\pi^2}M^2E_0\langle [\bar{q}
\{\gamma_{0} iD_{0}\} q]_1
\rangle_{\rho,I}L^{-\frac{4}{9}}\nonumber\\
&-\frac{4}{3}f\frac{\mathcal{R}_{-}(m_q)}{\mathcal{R}_{+}(m_q)}
\langle\bar{q}q\rangle_{\textrm{vac}}\langle[\bar{q}q]_0 \rangle_p
L^{\frac{4}{9}}-\frac{1}{4\pi\alpha_s}\frac{M_N}{2}\left(-
[T^1_{~1}-T^2_{~1}]+\frac{1}{3} [T^3_{~1}-T^4_{~1}]\right)
L^{-\frac{4}{9}}\nonumber\\
&+\bar{E}_q\left\{\frac{2}{3\pi^2}\langle [\bar{q} \{\gamma_{0}
iD_{0} iD_{0}\} q]_1
\rangle_{p}L^{-\frac{4}{9}}-\frac{1}{18\pi^2}\langle [g_s
q^\dagger\sigma\cdot \mathcal{G}
q]_1\rangle_{p}L^{-\frac{4}{9}}\right\}.
\end{align}

\end{widetext}

\section{Borel transformation}
To emphasize the quasi nucleon pole, the phenomenological side and
the OPE side have to be Borel transformed. The transformation
changes the phenomenological side to have the following weighed
dispersion relation:
\begin{align}
\mathcal{B}[\Pi_i(q_0,\vert\vec{q}\vert)]=&~\frac{1}{2\pi
i}\int^{\omega_0} _{-\omega_0}
d\omega \phantom{1}W(\omega)\Delta\Pi_i(\omega,\vert\vec{q}\vert),\\
W(\omega)=&~(\omega-\bar{E}_q)e^{-\omega^2/M^2}\label{weight},
\end{align}
where $\bar{E}_q$ is the quasi hole pole which will be assigned to
satisfy Eq.~\eqref{pole}. The weighting function will de-emphasize
the contribution from the quasi hole, and the Borel transformation
suppress the continuum contribution. Using Eq.~\eqref{dis-eo}, the
OPE side of the sum rule can be obtained by taking the Borel
transformation of
$\Pi_i(q_0,\vert\vec{q}\vert)=\Pi_i^E(q_0^2,\vert\vec{q}\vert)-\bar{E}_q
\Pi_i^O(q_0^2,\vert\vec{q}\vert)$. Here, we define the differential
operator $\mathcal{B}$ for the Borel transformation of the OPE side
as
\begin{align}
\mathcal{B}[f(q_0^2,\vert\vec{q}\vert)]\equiv & \lim_{\substack{-q_0^2,n\rightarrow\infty\\-q_0^2/n=M^2}} \frac{(-q_0^2)^{n+1}}{n!}\left(\frac{\partial}{\partial
q_0^2}\right)^n f(q_0^2,\vert\vec{q}\vert)\nonumber\\
\equiv&~\hat{f}(M^2,\vert\vec{q}\vert),
\end{align}
where $M$ is the Borel mass~\cite{Reinders:1984sr}. Polynomial terms
in the OPE side vanish after the Borel transformation.

\end{document}